\documentclass[a4paper,twocolumn,superscriptaddress,pre, aps, floatfix ]{revtex4-1}
\usepackage{amssymb}
\usepackage{amsmath}
\usepackage{marvosym}
\usepackage{graphicx}

\usepackage{url}
\usepackage{lipsum}
\usepackage{soul}
\usepackage{multirow}
\usepackage[colorlinks=true,allcolors=blue]{hyperref}
\usepackage{booktabs}
\usepackage{latexsym}
\usepackage{amsbsy}

\usepackage{epstopdf}
\usepackage{verbatim}
\usepackage[usenames]{color}
\usepackage{capt-of} 

\usepackage{ esint }
\usepackage{hyperref}
\usepackage[normalem]{ulem}
\usepackage{setspace}
\usepackage{caption}

\usepackage{tabularx}
\usepackage{xr}
\usepackage{xcolor}
\usepackage{float}
\floatstyle{ruled} 
\newfloat{algorithm}{t}{lop} 
\usepackage{algorithm}
\usepackage{algpseudocode}  

\newfloat{algorithm}{tbp}{loa}
\floatname{algorithm}{Algorithm}

\usepackage{placeins} 
\FloatBarrier  
\usepackage[justification=justified,singlelinecheck=false]{caption} 
\usepackage{ragged2e} 

\begin{document}

\title{Clustering Does Not Always Imply Latent Geometry}

\author{Roya Aliakbarisani}
\email{roya\_aliakbarisani@ub.edu}
\affiliation{Departament de F\'isica de la Mat\`eria Condensada, Universitat de Barcelona, Mart\'i i Franqu\`es 1, E-08028 Barcelona, Spain}
\affiliation{Universitat de Barcelona Institute of Complex Systems (UBICS), Barcelona, Spain}

\author{Mari\'an Bogu\~n\'a}
\email{marian.boguna@ub.edu}
\affiliation{Departament de F\'isica de la Mat\`eria Condensada, Universitat de Barcelona, Mart\'i i Franqu\`es 1, E-08028 Barcelona, Spain}
\affiliation{Universitat de Barcelona Institute of Complex Systems (UBICS), Barcelona, Spain}

\author{M. \'Angeles Serrano}
\email{marian.serrano@ub.edu}
\affiliation{Departament de F\'isica de la Mat\`eria Condensada, Universitat de Barcelona, Mart\'i i Franqu\`es 1, E-08028 Barcelona, Spain}
\affiliation{Universitat de Barcelona Institute of Complex Systems (UBICS), Barcelona, Spain}
\affiliation{Instituci\'o Catalana de Recerca i Estudis Avan\c{c}ats (ICREA), Passeig Llu\'is Companys 23, E-08010 Barcelona, Spain}

\begin{abstract}
The latent space approach to complex networks has revealed fundamental principles and symmetries, enabling geometric methods. However, the conditions under which network topology implies geometricity remain unclear. We provide a mathematical proof and empirical evidence showing that the multiscale self-similarity of complex networks is a crucial factor in implying latent geometry. Using degree-thresholding renormalization, we prove that any random scale-free graph in a $d$-dimensional homogeneous and isotropic manifold is self-similar when interactions are pairwise. Hence, both clustering and self-similarity are required to imply geometricity. Our findings highlight that correlated links can lead to finite clustering without self-similarity, and therefore without inherent latent geometry. The implications are significant for network mapping and ensemble equivalence between graphs and continuous spaces.
\end{abstract}

\maketitle
Network geometry~\cite{Boguna2021} has emerged as a key paradigm in network science to model real-world networks at both local and global scales. Moreover, it provides a robust framework to conceptualize problems in physics related to ensemble equivalence between graphs of discrete units and emerging continuous space or spacetimes, including approaches to quantum gravity~\cite{Wu:2015et} and the study of causal sets~\cite{Bombelli:1987im,Surya:2019aa,Krioukov:2012qf}. In particular, the latent geometry approach~\cite{Penrose:2003ud}---where nodes connect with a likelihood that decreases with distance in a hidden metric space---explains essential network features~\cite{serrano_boguna_2022}. Notably, this approach has revealed that real networks exhibit an effective hyperbolic geometry~\cite{Krioukov2010,Candellero03032016}, unifying the small-world property with heterogeneous degree distributions and clustering under a single mechanism. 

The key property that connects the hyperbolic geometry of a network and its topology is clustering, the tendency of nodes to share neighbors~\cite{newman2003properties}, with triangles in the network structure induced as a reflection of the triangle inequality in the metric space. This raises the intriguing question of whether clustering implies geometricity. Previous work~\cite{PhysRevLett.116.208302} suggested that clustering implies geometry by showing that random graphs with homogeneous degrees and clustering are equivalent to random geometric graphs. However, this holds only under certain conditions. Furthermore, network models without any inherent geometry, like the configuration model, can still contain large amounts of triangles. The limitations of using triangle counts to detect geometry led to the development of a weighting scheme designed to determine reliably whether clustering is induced by hyperbolic spaces and persists in the thermodynamic limit, which requires system size scaling~\cite{PhysRevE.106.054303}. 

In this work, we unveil the crucial role of network symmetries to elucidate the conditions under which network topology implies geometricity. We used the degree-threhsolding renormalization (DTR) method~\cite{Serrano2008}, which generates a nested hierarchy of subgraphs by progressively filtering out nodes with degree below a threshold, to reveal multiscale self-similarity. DTR self-similarity has been observed in many real-world networks and has been explained by the renormalizability of the geometric $\mathbb{S}^1$ model~\cite{Serrano2008}. Self-similarity of complex networks has also been explored from other perspectives~\cite{Song2005, Song2006, Kim2007, alvarez-hamelin2008, garcia2018multiscale}. 

\begin{figure}[h!]
\centering
\includegraphics[width=0.5\textwidth]{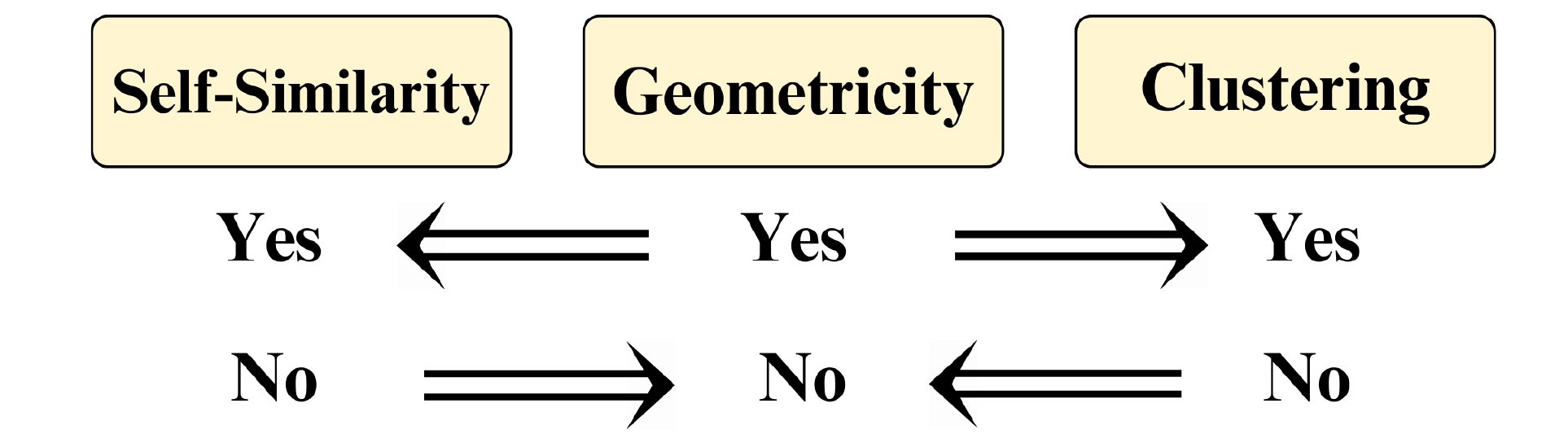}
\caption{Logic consequence relationships between geometricity, clustering, and self-similarity for geometric random graphs in homogeneous and isotropic spaces.}

\label{Fig:GSC-Relationship}
\end{figure}

Below, {\it we prove that any geometric random graph in a homogeneous and isotropic manifold with any curvature and any dimension is self-similar under DTR, provided nodes have a scale-free degree distribution and pairwise interactions. Therefore, non-self-similar networks cannot be geometric.  We found that some models with finite clustering are non-self-similar under DTR, thus non-geometric.} Hence, while geometricity always implies clustering and self-similarity, the reverse is not true; however, the absence of clustering or self-similarity implies non-geometricity, see Fig.~(\ref{Fig:GSC-Relationship}). Moreover, real networks can be classified according to their self-similarity and clustering levels, providing insights into their potential geometricity. Thus, clustering, geometry, and self-similarity interplay in a non-trivial way in complex networks.

We start our proof by considering the thermodynamic limit of a homogeneous Poisson point process with density $\delta$ in a Euclidean space of dimension $d$. The same is obtained for homogeneous and isotropic hyperbolic and spherical manifolds; the proofs for negative and positive curvature are provided in Appendices~\ref{sec:RGG-hyperbolic} and ~\ref{sec:RGG-spherical}, respectively. Note that the proof for spherical manifolds applies, in particular, to the $\mathbb{S}^d$ geometric network model~\cite{Serrano2008} and, thus, to its hyperbolic formulation the $\mathbb{H}^{d+1}$ model~\cite{jankowski2023d,budel2024random}, where homogeneous nodes are distributed heterogeneously in the $(d+1)$-dimensional hyperbolic space. In contrast, the proof for hyperbolic manifolds refers to heterogeneous nodes distributed homogeneously over all the $d$-dimensional hyperbolic space. 

Such a space conceptualizes a similarity space, in which closeness reflects how similar nodes are and determines their likelihood to form connections. Furthermore, we assume that each node has a popularity attribute, independent of its position in the space, ultimately accounting for its number of neighbors in the network. We represent popularity by a hidden variable $h$ (a positive real number), distributed as $\rho(h)$. Suppose now that interactions are pairwise, such that the connection probability between a pair of nodes $i$ and $j$ with hidden variables $h_i$ and $h_j$ and separated by a distance $x_{ij}$ takes the form
\begin{equation}
p_{ij}=f\left(\frac{x_{ij}}{h_i h_j}\right).
\end{equation}
We do not impose any particular form to the function $f(\cdot)$ other than being integrable in $\mathbb{R}^{d}$. As we show below, this implies that the model generates sparse networks without the need to introduce a size dependence on the connection probability. Since clustering is defined as the probability that two nodes that share a common neighbor are connected, it is a function of $p_{ij}$ and must be finite in the thermodynamic limit. This defines our ensemble of random geometric graphs $\mathcal{G}$.

In this setting, the expected degree of node $i$ is simply given by $\bar{k}(h_i)=\sum_{j \ne i} p_{ij}$. However, since the space is homogeneous and isotropic, and the hidden variable $h$ is independent of space, we can place node $i$ at the origin of coordinates so that $x_{ij}$ becomes the radial distance between the nodes in spherical coordinates. Then, taking the continuum limit,
\begin{equation}
\bar{k}(h_i)= S_{d-1} \delta \int dh_j \rho(h_j) \int dx_{ij} x_{ij}^{d-1}f\left(\frac{x_{ij}}{h_i h_j}\right),
\end{equation}
where we have already integrated the angular part and $S_{d-1}$ is the volume of the $(d-1)$-sphere. A simple change of variables leads to $\bar{k}(h_i)=\delta I_d \langle h^d \rangle h_i^d$,
where $I_d \equiv S_{d-1}\int_0^{\infty} f(x) x^{d-1}dx$, so that the average degree is 
\begin{equation}
\langle k \rangle=\delta I_d\langle h^d \rangle^2.
\label{eq:average_T}
\end{equation}
As discussed earlier, the average degree is constant --and so the network is sparse-- as long as the integral $I_d$ is bounded. Hereafter, we constraint the distribution of hidden variables $h$ to be scale-free with $\rho(h) = (\tilde{\gamma} - 1) h_0^{(\tilde{\gamma} - 1)} h^{-\tilde{\gamma}}$, $h \geq h_0$ and $\tilde{\gamma}>d$.

We are now interested in the properties of subgraphs obtained after applying the DTR procedure to networks generated by the ensemble $\mathcal{G}$. Given a graph $G \in \mathcal{G}$, we obtain a subgraph $G_T$ by removing all nodes with hidden variable $h_0 \le h \le h_T$ from $G$, defining the ensemble $\mathcal{G}_T$. Since we are truncating a power-law distribution from below, the hidden variables of nodes in $G_T$ are also power-law distributed with the same exponent $\tilde{\gamma}$ but starting at $h_T$ instead of $h_0$. Hence, $\delta_T / \delta= \lim_{N\rightarrow \infty}N_T / N$, where $N$ and $N_T$ are the sizes of the original network and the subgraph. Since the popularity dimension encoded in the hidden variable $h$ is uncorrelated with the position in the metric space, the ensemble $\mathcal{G}_T$ is the same as $\mathcal{G}$ in the thermodynamic limit, with the only difference that the density of the Poisson point process $\delta_T$ and the average $\langle h^d \rangle_T$ are rescaled. Hence, the ensemble $\mathcal{G}$ is statistically invariant under DTR and it contains an infinite hierarchy of nested subgraphs belonging to the same ensemble. The only relevant observable is the average degree, which may change in the DTR flow. Its renormalized value is obtained from Eq.~\eqref{eq:average_T}, and reads
\begin{equation}
\langle k \rangle_T = \langle k \rangle \left(\frac{\delta}{\delta_T}\right)^{\frac{3 - \gamma}{\gamma - 1}},
\end{equation}
where we have used
\begin{equation}
\delta_T = \delta \left(\frac{h_0}{h_T}\right)^{(\tilde{\gamma} - 1)},\;  \; \langle h^d \rangle_T = \langle h^d \rangle \left(\frac{h_T}{h_0}\right)^{d},
\end{equation}
and the definition of the expected degree of node $i$, which is proportional to $h^d$ as $\kappa_i \equiv \langle k \rangle h_i^d / \langle h^d \rangle$. This gives $\langle \kappa \rangle = \langle k \rangle$, and $\kappa$ power-law distributed with an exponent $\gamma = 1 + (\tilde{\gamma} - 1)/d$. Since $\kappa$ is a monotonically increasing function of $h$, thresholding by $h_T$ is equivalent to thresholding by $\kappa_T$. 

We can analyze the DTR flow of the average degree as a function of the relative size of $G_T$ with respect to the size of $G$. Notice that it is independent of the dimension. For scale-free networks with $2 < \gamma < 3$, the average degree of subgraphs $G_T$ grows as one goes deeper into the hierarchy of subgraphs, while for $ \gamma = 3$ it remains stable, and for $ \gamma >3 $ it decreases, independently of the dimension.

\begin{figure*}[t!]
\centering
\includegraphics[width=1\textwidth]{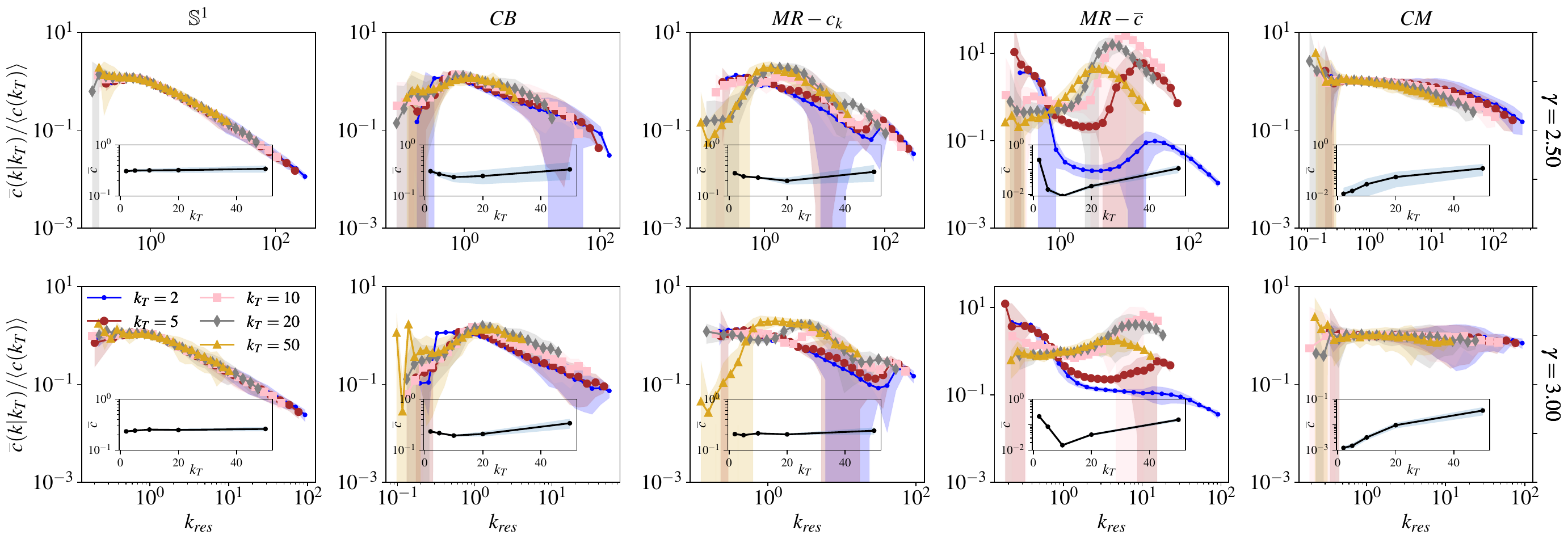}
\caption{Rescaled clustering coefficient as a function of rescaled degree in networks with $\beta = 1.5$, $\langle k \rangle = 10$ and $N = 50,000$. The insets illustrate the average clustering coefficient as a function of $k_T$. The shaded areas represent two-$\sigma$ intervals around the mean, calculated from 10 realizations of each model. The values of $\epsilon^2_{\overline{c}}$ for each model are: $0.08$ for $\mathbb{S}^1$, $0.87$ for the CB model, $6.05$ for $MR-c_k$, $5.76$ for $MR-\overline{c}$, and $0.16$ for the CM model.
 }\label{Fig:Clustering}
\end{figure*}

Hence, we have demonstrated that a geometric random graph with pairwise interactions and a scale-free degree distribution defined in a Euclidean metric space of any dimension is necessarily self-similar under DTR and the only property that may change in the flow is the average degree. Therefore, if a network does not exhibit self-similarity across the DTR hierarchy of subgraphs, it cannot be geometric.

We analyzed the DTR self-similarity of various network models with different clustering profiles to understand their geometricity. The models considered are the Configuration Model (CM)~\cite{BENDER1978296,Molloy:1995bf}, the geometric soft configuration model or $\mathbb{S}^1$ model~\cite{Serrano2008}, the clique-based model (CB)~\cite{Gleeson2009}, and two versions of the maximal randomization model (MR)~\cite{Colomer2013}, named $MR-c_k$ and $MR-\overline{c}$, applied to a seed network with a given degree distribution. The CM generates graphs with a fixed degree distribution and no clustering in large networks. The other models also have a fixed degree distribution but produce graphs with finite clustering in the thermodynamic limit. The $\mathbb{S}^1$ model maximizes entropy while maintaining a given degree distribution and clustering level~\cite{Boguna:2020fj}. The CB achieves clustering by producing disjoint cliques connected by extra links. The $MR-c_k$ and the $MR-\overline{c}$ tune the clustering spectrum and average clustering coefficient, respectively, by rewiring network connections to achieve target clustering levels. The CM and $\mathbb{S}^1$ are pairwise, meaning the connection likelihood between two nodes is independent of other nodes, while the CB, $MR-c_k$, and $MR-\overline{c}$ are not. See Table~\ref{tab:method_comparison}, and more details on the models in Appendix~\ref{sec:Random-graph-models}.

We generated synthetic networks with the described models and studied the change of their structural properties in the DTR flow. For each network, we applied the degree thesholding method to produce a nested hierarchy of subgraphs. Since all networks had the same degree distributions, the threshold values $k_T$ were fixed to predefined levels, $\{2, 5, 10, 20, 50 \}$, ensuring consistency in the analysis across different network models. For each subgraph, we measured the degree distribution, the average nearest-neighbors degree, and the clustering coefficient as functions of the degrees rescaled by the average degree of the corresponding subgraph $k_{res}=k/\left< k(k_T)\right>$. Multiscale self-similarity is denoted by overlapping curves for the different subgraphs in the DTR flow, while spread curves denote scale-dependent behavior. 

Fig.~\ref{Fig:Clustering} shows the behavior of the clustering coefficient. The $\mathbb{S}^1$ and $CM$ models display self-similarity, both $MR$ models and the CM are clearly non-self-similar, and the CB model shows an intermediate behavior that could be named quasi-self-similar. According to our proof, the lack of self-similarity of $MR-c_k$ and $MR-\overline{c}$ implies that they lack a latent geometry, even if they generate high clustering. Interestingly, the peculiar behavior of $MR-\overline{c}$ is indicative of a core-periphery structure and double percolation phenomena explored in~\cite{Colomer-de-Simon:2014yo}.  A detailed analysis of synthetic networks for additional parameter values for $\beta$ and $\gamma$ are in Appendix.~\ref{sec:exp-synthetic}, (Figs.~\ref{fig:Synthetic_clustering_b_1.5}, \ref{fig:Synthetic_clustering_b_3}), where we also report other structural properties, including the complementary cumulative degree distribution (Figs.~\ref{fig:Synthetic_CCD_b_1.5}, \ref{fig:Synthetic_CCD_b_3}) and the average nearest-neighbor degree function (Figs~\ref{fig:Synthetic_Knn_b_1.5}, \ref{fig:Synthetic_Knn_b_3}) which display congruent behaviors. Our conclusions about the self-similarity and geometricity of the models are summarized in Table~\ref{tab:method_comparison}.

\begin{table}[ht]
    \centering
    \caption{Properties of network models.}
    \resizebox{\columnwidth}{!}{ 
        \begin{tabular}{lcccc} 
            \toprule
            \hline
            \textbf{Model}  & \textbf{Pairwise} & \textbf{Clustering}  & \textbf{Self-Similarity} & \textbf{Geometric} \\
            \midrule
            \hline
            \hline
            $\mathbb{S}^1$ & Yes & Yes & Yes & Yes \\
            $CB $ & No & Yes  & No & No \\
            $MR-c_{k}$ & No & Yes  & No & No \\
            $MR-\overline{c}$ & No & Yes  & No & No \\
            $CM$  & Yes & No  & Yes & No \\
            \hline
            \hline
            \bottomrule
        \end{tabular}
    }
    
    \label{tab:method_comparison}
\end{table}

We also analyzed real-world networks using the same methodology. Their description and structural properties are reported in Appendix~\ref{sec:Data-description} and Table~\ref{tab:network_properties}. Due to finite-size effects and varying degree distributions, values for $k_T$ were chosen in $[2, k_T^{\max}]$ for every network in enough number to ensure adequate resolution, where the maximum $k_T^{\max}$ was determined by taking the maximum value of the average degree in the DTR flow (see Fig.~\ref{fig:Avg_Degree_vs_KT} in Appendix~\ref{sec:epsilon-test}). In Fig.~\ref{Fig:Real_Networks}-(a), we display the DTR flow of the structural properties of two real networks. The curves of Fb-Friends (top row) overlap, suggesting similar structure of the renormalized networks. In contrast, for PPI-rat (bottom row), the structural properties vary across different subgraphs. In Appendix~\ref{sec:Exp-Real}, we show the structural properties of all the real-world networks analyzed here, classified as self-similar (Figs.~\ref{fig:Real_Net_Self_Similar1}, \ref{fig:Real_Net_Self_Similar2}, \ref{fig:Real_Net_Self_Similar3}), quasi-self-similar (Fig.~\ref{fig:Real_Net_Quasi_Self_Similar}) and non-self-similar (Fig.~\ref{fig:Real_Net_Non_Self_Similar}). 

\begin{figure}[ht]
\centering
\includegraphics[width=\columnwidth]{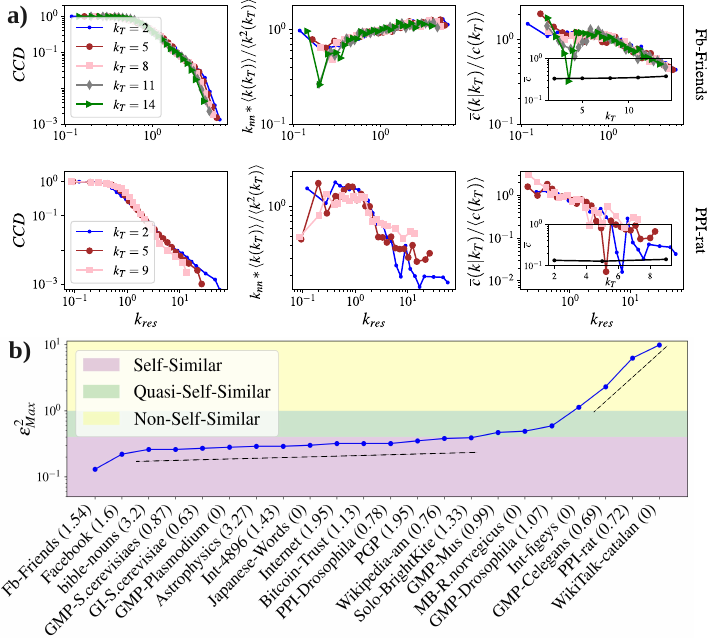}
\caption{Geometricity of real networks. (a) Complementary cumulative degree distribution, average nearest-neighbors’ degree, and clustering spectrum as functions of the rescaled degree for a self-similar network (top) and a non-self-similar (bottom) network. Insets show the average clustering vs. $k_T$. (b) Distribution of $\epsilon^2_{\text{Max}}$ values for real networks, sorted in ascending order. The three classification regions are: self-similar (purple, $\epsilon^2_{\text{Max}} < 0.4$), quasi-self-similar (green, $0.4 \leq \epsilon^2_{\text{Max}} < 1$), and non-self-similar (yellow, $\epsilon^2_{\text{Max}} \geq 1$). Black dashed lines highlight trends. The value of $\beta$ of every network is displayed in parenthesis. 
}
\label{Fig:Real_Networks}
\end{figure}

Fig.~\ref{Fig:Real_Networks}-(b) displays the classification of real networks as self-similar, quasi-self-similar, or non-self-similar, see also the Sankey diagram in Fig.~\ref{fig:Sankey} of Appendix~\ref{sec:Exp-Real}. This classification is based on an $\epsilon^2$-Test, which calculates the maximum difference $\epsilon^2_{\text{Max}}$ between the curves of the original network and its subgraphs in the range of rescaled degrees common to both, computed as  $\epsilon^2_{\text{Max}} = Max(\epsilon^2_{CCD}, \epsilon^2_{knn} , \epsilon^2_{\overline{c}})$. The values for the the three structural properties were obtained by averaging the squared relative differences between the corresponding curves of the original network and the subgraphs. For more details, refer to the algorithm in Appendix~\ref{sec:epsilon-test}. In Fig.~\ref{Fig:Real_Networks}-(b), networks are arranged in ascending order along the x-axis based by increasing $\epsilon^2_{\text{Max}}$. The curve shows three distinct regions: low values represent self-similar networks, which exhibit a gradual increase with a very smooth slope, a moderate increase marks quasi-self-similar networks, and a sharp rise indicates non-self-similar networks. Based on this, networks are categorized as self-similar ($\epsilon^2_{\text{Max}} < 0.4$), quasi-self-similar ($0.4 \leq \epsilon^2_{\text{Max}} < 1$), and non-self-similar ($\epsilon^2_{\text{Max}} \geq 1$). According to these bounds, Fb-Friends ($\epsilon^2_{\text{Max}} = 0.12$) is self-similar, while PPI-rat ($\epsilon^2_{\text{Max}} = 6.27$) is non-self-similar. This scheme is also applicable to synthetic networks, correctly classifying $\mathbb{S}^1$ and $CM$ as self-similar, $CB$ as quasi-self-similar, and the two versions of the $MR$-Model as non-self-similar, see values reported in the caption of Fig.~\ref{Fig:Clustering}.

For a consistency test, we calculated the level of coupling between the topology of the networks and the hyperbolic two-dimensional latent space of the $\mathbb{S}^1$ model. This coupling is quantified by the parameter $\beta$, which controls the level of clustering in the model and can be estimated using an extended version of the Mercator embedding tool capable of handling both strongly and weakly geometric networks~\cite{van_der_Kolk2022}. The embedding adjusts the coordinates of nodes in the latent space and infers the parameters such that the observed topology of the network is best reproduced by synthetic networks of the model. 

The $\beta$ values for real networks, shown in Fig.~\ref{Fig:Real_Networks}-(b) and Fig.~\ref{fig:Sankey}, indicate that the absence of DTR self-similarity implies non-geometricity. The results reveal that non-self-similar networks have $\beta < 1$, categorizing them as either non-geometric ($\beta \lesssim 0.5$) or quasi-geometric ($0.5 \lesssim \beta < 1$). Geometric networks ( $\beta > 1$) are self-similar, except for GMP-Drosophila, which is quasi-self-similar with $\beta$ near the critical value $\beta_c = 1$. The average $\beta$ values of self-similar, quasi-self-similar, and non-self-similar networks are $1.36$, $0.69$, and $0.35$, respectively. Table~\ref{tab:network_properties} in Appendix~\ref{sec:Exp-Real} shows that geometric networks with $\beta > 1$ have high clustering coefficients. However, as noted in~\cite{van_der_Kolk2022}, networks in the quasi-geometric and non-geometric regions, where topology couples weakly wtih the underlying metric space, exhibit clustering coefficients that tends to zero in the thermodynamic limit. Despite this, clustering decay is slow, especially in the quasi-geometric regime, so finite systems may still retain significant clustering. In conclusion, networks that fit a geometric description are self-similar and highly clustered. However, self-similar networks are not always geometric, and non-self-similar networks tend to be weakly geometric or incompatible with a geometric description.

To conclude, we used DTR to elucidate the relation between clustering, geometry, and self-similarity. We demonstrated that any geometric random graph in a homogeneous and isotropic manifold (with zero, positive, or negative curvature) of any dimension is self-similar under DTR, when nodes have a scale-free degree distribution and the interactions are pairwise. Therefore, non-self-similar networks cannot be geometric. While geometricity always implies both clustering and self-similarity, neither alone is sufficient to imply geometricity. We found that some network models with finite clustering are non-self-similar under DTR and must therefore be non-geometric. These findings seem to contradict~\cite{PhysRevLett.116.208302}. However, this contradiction is only apparent, highlighting that clustering implies latent geometry only when interactions are pairwise independent. Correlated links can result in finite clustering without inherent latent geometricity. Moreover, real networks can be classified by their self-similarity and clustering, which allows us assess their geometricity when both properties are present. 

Our work unveils how the existence of multiscale symmetries in networks helps in delimiting network geometry. If the conditions are met, a myriad of downstream tasks based on network maps becomes possible and meaningful for networks. The implications are far-reaching and important for any problem related to ensemble equivalence between graphs and continuous spaces.

\section*{Acknowledgments}
We acknowledge support from: Grant TED2021-129791B-I00 funded by MCIN/AEI/10.13039/501100011033 and by  ``European Union NextGenerationEU/PRTR''; Grant PID2022-137505NB-C22 funded by MCIN/AEI/10.13039/501100011033 and by ``ERDF A way of making Europe''; Generalitat de Catalunya grant number 2021SGR00856. M. B. acknowledges the ICREA Academia award, funded by the Generalitat de Catalunya. 

\clearpage
\onecolumngrid
\appendix

\section{Proof of the self-similarity of random geometric graphs in hyperbolic geometry} \label{sec:RGG-hyperbolic}

Let us consider the $d$-dimensional hyperbolic space with radius of curvature $R$, $\mathbb{H}^d$, where the square of the line element is defined as
\begin{equation}
ds^2 = dx^2 + R^2 \sinh^2\left(\frac{x}{R}\right) d\Omega_{d-1}^2,
\end{equation}
and  $d\Omega_{d-1}^2$ is the squared line element of a $(d-1)$-sphere of unit radius. Thus, $x \in (0,\infty)$ is the hyperbolic distance between a point at the origin of coordinates and the set of points with fixed angular coordinates on the $(d-1)$-sphere. Therefore, the volume element is
\begin{equation}
dV = R^{d-1} \sinh^{d-1}\left(\frac{x}{R}\right) dx \, d\Omega_{d-1}.
\end{equation}
We now define a Poisson point process on $\mathbb{H}^d$ at density $\delta$. This implies that the expected number of points within the volume element is simply $\delta dV$. Using this Poisson point process, we define a random geometric graph in the following way. We identify each point as a node in the graph and endow each node with a hidden variable $h$ drawn from the distribution $\rho(h)$, independently of the position of the node in $\mathbb{H}^d$. Then, a pair of nodes with hidden variables $h$ and $h'$, separated by the hyperbolic distance $x$, gets connected with probability
\begin{equation}
p = f\left(\frac{1}{hh'}\int_0^{\frac{x}{R}} \sinh^{d-1}{z} dz \right),
\label{connectionprob}
\end{equation}
where $0 \le f(\cdot) \le 1$ is an arbitrary but integrable function of its argument. Notice that the argument of function $f$ is a monotonically increasing function of the hyperbolic distance. Thus, Eq.~\eqref{connectionprob} represents the most general class of heterogeneous random geometric graphs with pairwise interactions in hyperbolic geometry .

Since $\mathbb{H}^d$ is a homogeneous and isotropic space, all nodes in the graph are geometrically equivalent and are only distinguished by their hidden variables $h$. Let us then consider a node at the origin of coordinates with hidden variable $h$ and compute its expected degree $\bar{k}(h)$ as
\begin{equation}
\bar{k}(h) = \delta R^{d-1} S_{d-1} \int \rho(h') dh' \int_0^\infty \sinh^{d-1}\left(\frac{x}{R}\right) f\left(\frac{1}{hh'}\int_0^{\frac{x}{R}} \sinh^{d-1}{z} dz \right) dx,
\end{equation}
where $S_{d-1}$ is the volume of the $(d-1)$-sphere. By making the change of variables
\begin{equation}
t = \frac{1}{hh'}\int_0^{\frac{x}{R}} \sinh^{d-1}{z} dz,
\end{equation}
we obtain
\begin{equation}
\bar{k}(h) = \delta R^{d} S_{d-1} I \langle h \rangle h, \quad \text{with} \quad I \equiv  \int_0^\infty  f\left(t \right) dt.
\end{equation}
We then see that the hidden variable $h$ is proportional to the expected degree of the node, so the degree distribution is determined by the distribution $\rho(h)$. The average degree is then
\begin{equation}
\langle k \rangle = \delta R^{d} S_{d-1} I \langle h \rangle^2.
\end{equation}
We are interested in scale-free networks. Thus, we choose $h$ to be power-law distributed as 
\[
\rho(h) = (\gamma-1) h_0^{\gamma-1} h^{-\gamma}, \quad \text{with} \quad h \ge h_0,
\]
so that $\langle h \rangle = (\gamma-1) h_0 / (\gamma-2)$.

Now, we apply the DTR transformation to the hidden variable and decimate the network by removing all nodes with hidden variables in the range $h_0 < h < h_T$. By doing so, the remaining nodes are also power-law distributed, but with hidden variables starting at $h_T$ instead of $h_0$, so that their average is
\begin{equation}
\langle h \rangle_T = \frac{h_T}{h_0} \langle h \rangle.
\label{eq:hmean}
\end{equation}
Furthermore, the relative size of the network after decimation compared to the original network is simply given by
\begin{equation}
\frac{N_T}{N} = \frac{\delta_T}{\delta} = \left(\frac{h_0}{h_T}\right)^{\gamma-1}.
\label{eq:NT}
\end{equation}
Therefore, the subgraph obtained after decimation is a Poisson point process on $\mathbb{H}^d$ at density $\delta_T$ and with hidden variables power-law distributed with exponent $\gamma$ and average $\langle h \rangle_T$. The subgraph is then a realization of the same ensemble as the original network, with the only difference being that the average degree is given by
\begin{equation}
\langle k \rangle_T = \langle k \rangle \left(\frac{N}{N_T} \right)^\frac{3-\gamma}{\gamma-1},
\end{equation}
just as in the Euclidean case.

\section{Proof of the self-similarity of random geometric graphs in spherical geometry} \label{sec:RGG-spherical}

We now turn to geometries with constant positive curvature. Let us consider the $d$-dimensional spherical space with radius of curvature $R$, $\mathbb{S}^d$, where the square of the line element is 
\begin{equation}
ds^2 = dx^2 + R^2 \sin^2\left(\frac{x}{R}\right) d\Omega_{d-1}^2,
\end{equation}
and $d\Omega_{d-1}^2$ is the squared line element of a $(d-1)$-sphere of unit radius. Thus, $x \in (0,\pi R)$ is the spherical distance between the origin of coordinates and the set of points with fixed angular coordinates on the $(d-1)$-sphere. Therefore, the volume element is
\begin{equation}
dV = R^{d-1} \sin^{d-1}\left(\frac{x}{R}\right) dx \, d\Omega_{d-1}.
\end{equation}
A key difference between spherical and hyperbolic geometries is that the former are compact manifolds with a finite volume given by
\begin{equation}
V_{\text{tot}}=\frac{2 \pi^{\frac{d+1}{2}}}{\Gamma\left(\frac{d+1}{2} \right)}R^d.
\end{equation}
Thus, if we define a Poisson point process on $\mathbb{S}^d$ at density $\delta$, the expected number of nodes in the graph is given by
\begin{equation}
N=\frac{2 \pi^{\frac{d+1}{2}}}{\Gamma\left(\frac{d+1}{2} \right)}\delta R^d.
\end{equation}
This means that, to take the thermodynamic limit, we can either let $R \gg 1$ while keeping $\delta$ constant or, alternatively, keep $R$ fixed and let $\delta \gg 1$ or, in fact, any other combination that leads to $\delta R^d \gg 1$. As in the hyperbolic case, we endow each node with a hidden variable $h$ drawn from the distribution $\rho(h)$, independently of the node's position in $\mathbb{S}^d$. Then, a pair of nodes with hidden variables $h$ and $h'$, separated by the spherical distance $x$, are connected with probability
\begin{equation}
p = f\left(\frac{\delta R^d}{hh'}\int_0^{{\frac{x}{R}}} \sin^{d-1}{z} \, dz \right),
\label{connectionprobSd}
\end{equation}
where $f(\cdot)$ is an arbitrary but integrable function. Notice that, contrary to the hyperbolic case, we include an explicit dependence in the connection probability on the term $\delta R^d$, which is proportional to the system size. This choice is taken so that the model defines an ensemble of sparse networks in the thermodynamic limit. Since function $f(\cdot)$ is any integrable function of its argument, Eq.~\eqref{connectionprobSd} represents the most general class of heterogeneous random geometric sparse graphs with pairwise interactions on $\mathbb{S}^d$.

Since $\mathbb{S}^d$ is a homogeneous and isotropic space, all nodes in the graph are geometrically equivalent and differ only by their hidden variables $h$. Let us then consider a node at the origin with hidden variable $h$ and compute its expected degree $\bar{k}(h)$ as
\begin{equation}
\bar{k}(h) = \delta R^{d-1} S_{d-1} \int \rho(h') \, dh' \int_0^{\pi R} \sin^{d-1}\left(\frac{x}{R}\right) f\left(\frac{\delta R^d}{hh'}\int_0^{\frac{x}{R}} \sin^{d-1}{z} \, dz \right) dx.
\end{equation}
By making the change of variables
\begin{equation}
t = \frac{\delta R^d}{hh'}\int_0^{\frac{x}{R}} \sin^{d-1}{z} \, dz,
\end{equation}
we obtain
\begin{equation}
\bar{k}(h) = S_{d-1}\,h \int \rho(h')\,h' \, I(h,h'), 
\quad \text{with} \quad 
I(h,h') \equiv  \int_0^\frac{\alpha_d\delta R^d }{hh'}  f\bigl(t\bigr)\, dt,
\end{equation}
and $\alpha_d=$. In the limit of very large networks, $\delta R^d \gg 1$, and so the integral $I(h,h')$ becomes a constant. Thus, in this limit we can write
\begin{equation}
\bar{k}(h)  \approx S_{d-1}\,h\,I\,\langle h \rangle, 
\quad \text{with} \quad 
I \equiv  \int_0^\infty  f\bigl(t\bigr)\, dt.
\end{equation}
Again, we see that $h$ is simply proportional to the expected degree and $\langle k \rangle=S_{d-1}\,\,I\,\langle h \rangle^2$. As  in the  hyperbolic case, we are interested in scale-free networks, so that we consider $h>h_0$ to be power-law distributed with exponent $\gamma$.

Now, we apply the DTR transformation to the hidden variable and decimate the network by removing all nodes with hidden variables in the range $h_0 < h < h_T$. By doing so, the remaining nodes also follow a power-law distribution, but with hidden variables starting at $h_T$ instead of $h_0$, such that their average is given by Eq.~\eqref{eq:hmean}. Similarly, the relation between the size and density of the subgraph and the corresponding quantities of the original network is also given by Eq.~\eqref{eq:NT}. The connection probability between nodes in the subgraph remains the same as in the original network, thus retaining its explicit dependence on $\delta$. However, the density of nodes in the subgraph is $\delta_T$. Therefore, the average degree in the subgraph is given by
\begin{equation}
\langle k \rangle_T=\frac{\delta_T}{\delta}S_{d-1}\,\,I\,\langle h \rangle_T^2.
\end{equation}
Combining this result with Eqs.~\eqref{eq:hmean} and~\eqref{eq:NT}, we obtain 
\begin{equation}
\langle k \rangle_T = \langle k \rangle \left(\frac{N}{N_T} \right)^\frac{3-\gamma}{\gamma-1},
\end{equation}
which is, once again, the same as in the Euclidean case.

\section{Random Graph Models:} \label{sec:Random-graph-models}
\subsection{Clique-based Model (CB):}  \label{subsec:CB-Model}
The clique-based model (CB)~\cite{Gleeson2009}, also referred to as the Gleeson model, is a network model designed to replicate both the degree distribution and clustering properties commonly found in real-world networks. The CB model is governed by a joint probability distribution, $\gamma(k,c)$, which indicates the probability that a randomly chosen node has degree $k$ and belongs to a clique of size $c$, with $k \ge c-1$. Here, $k-c+1$ represents the number of external links that are connected to nodes outside the clique. After assigning each node to a single $c$-clique, nodes within each clique form super-nodes, and their external link stubs are randomly connected together using the configuration model. This approach allows for both local clustering and long-range interactions across the network.

The CB model can closely approximate the structural characteristics of real-world networks. However, as it starts from cliques rather than individual nodes, the number of nodes and their degrees may not be strictly predetermined, leading to small discrepancies between the expected and actual properties in finite networks.
Despite these variations, the CB model remains a valuable tool for simulating networks with high clustering and heterogeneous degree distributions.

\subsection{Maximally Random model (MR model):}  \label{subsec:MR-Model}
The Maximally Random model (MR)~\cite{Colomer2013} is a network model designed to realistically capture clustering while imposing minimal constraints. The MR model preserves the degree distribution and clustering coefficient for each degree level MR-$c_k$ or the average clustering coefficient of the entire network MR-$\overline{c}$ by employing an optimization approach based on a specific Hamiltonian. This Hamiltonian function measures the difference between the model's clustering and a target clustering coefficient, either at each degree level or across the whole network, allowing the model to control clustering levels while maintaining network randomness. A Metropolis-Hastings rewiring process and simulated annealing are used to minimize clustering deviations while keeping the degree sequence fixed. Consequently, the MR model generates clustered networks that closely resemble real-world networks in terms of clustering levels without enforcing an artificially ordered or modular structure.

\vspace{-10pt}
\subsection{$\mathbb{S}^1$ model:}  \label{subsec:S1-Model}
The $\mathbb{S}^1$ model~\cite{Serrano2008,serrano_boguna_2022} is a geometric model in a latent similarity metric space that generates synthetic networks with realistic properties, including high clustering~\cite{Boguna:2020fj, Krioukov2010, gugelmann2012random, candellero2016clustering, Fountoulakis2021}, heterogeneous degree distributions~\cite{Serrano2008, Krioukov2010, gugelmann2012random}, self-similarity~\cite{Serrano2008}, small-worldness~\cite{abdullah2017typical, friedrich2018diameter, muller2019diameter}, among others. 
In the $\mathbb{S}^1$ model, each node is characterized by two hidden variables, $(\kappa, \theta)$. The parameter $\kappa$ indicates the expected degree of the node, reflecting its popularity, while $\theta$ represents its angular position on a one-dimensional sphere, which abstracts the similarity space, with the radius proportional to the total number of nodes. The connection probability between two nodes follows a gravity-like law: nodes that are farther apart in similarity space are less likely to connect, whereas nodes with higher popularity are more likely to form connections.  
The parameter $\beta$, which is the inverse of temperature, controls the level of clustering. At the critical value $\beta_c = 1$, the system transitions from a high-clustering regime at high $\beta$ to a low-clustering regime at low $\beta$~\cite{Serrano2008}. 
It also regulates the coupling strength between the similarity space and the topology through the triangle inequality.

\section{Experimental Analysis of synthetic networks}  \label{sec:exp-synthetic}
In the following, we analyze the degree-thresholding renormalization (DTR) self-similarity of network models to assess their geometricity. Synthetic networks were initially generated using the $\mathbb{S}^1$ model with a total of $N=50000$ nodes and varying parameter values for $\beta$ and $\gamma$, while keeping the average degree fixed at $\langle k \rangle = 10$. These networks were subsequently randomized using random graph models, including the $CB$, $MR$, and $CM$ models.

\begin{figure*}[t!]
\centering
\includegraphics[width=\textwidth]{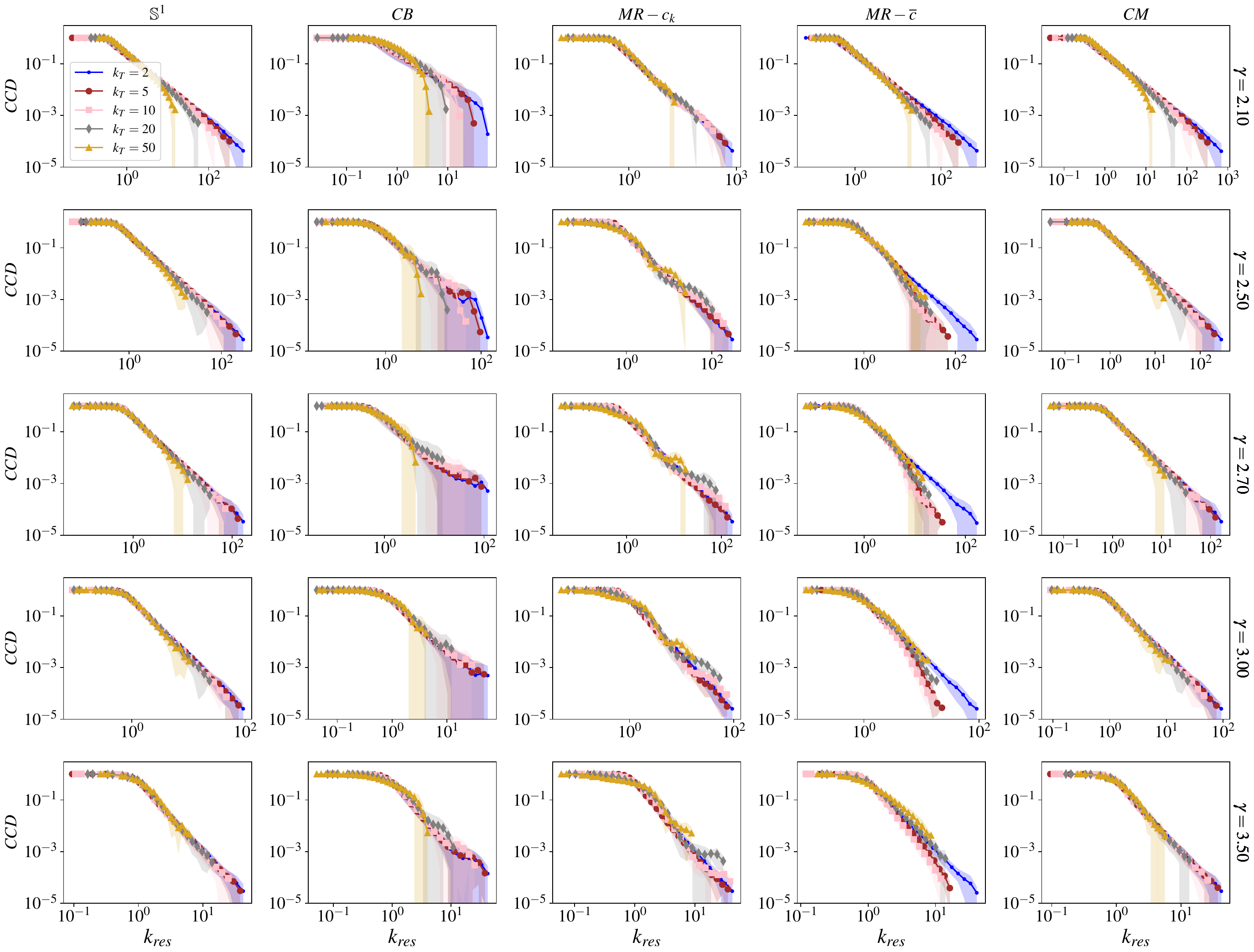}
\caption{The complementary cumulative degree distribution of the synthetic networks in the DTR flow as a function of the degree rescaled by the average degree in the corresponding subgraph, $k_{\text{res}} = k / \langle k(k_T) \rangle$. Each column corresponds to synthetic networks generated by a model with $N = 50000$ nodes, $\langle k \rangle = 10$, $\beta = 1.5$ and varying values of $\gamma$. The shaded areas represent two-$\sigma$ intervals around the mean, calculated from 10 realizations of each model. }\label{fig:Synthetic_CCD_b_1.5}
\end{figure*}

\begin{figure*}[t!]
\centering
\includegraphics[width=\textwidth]{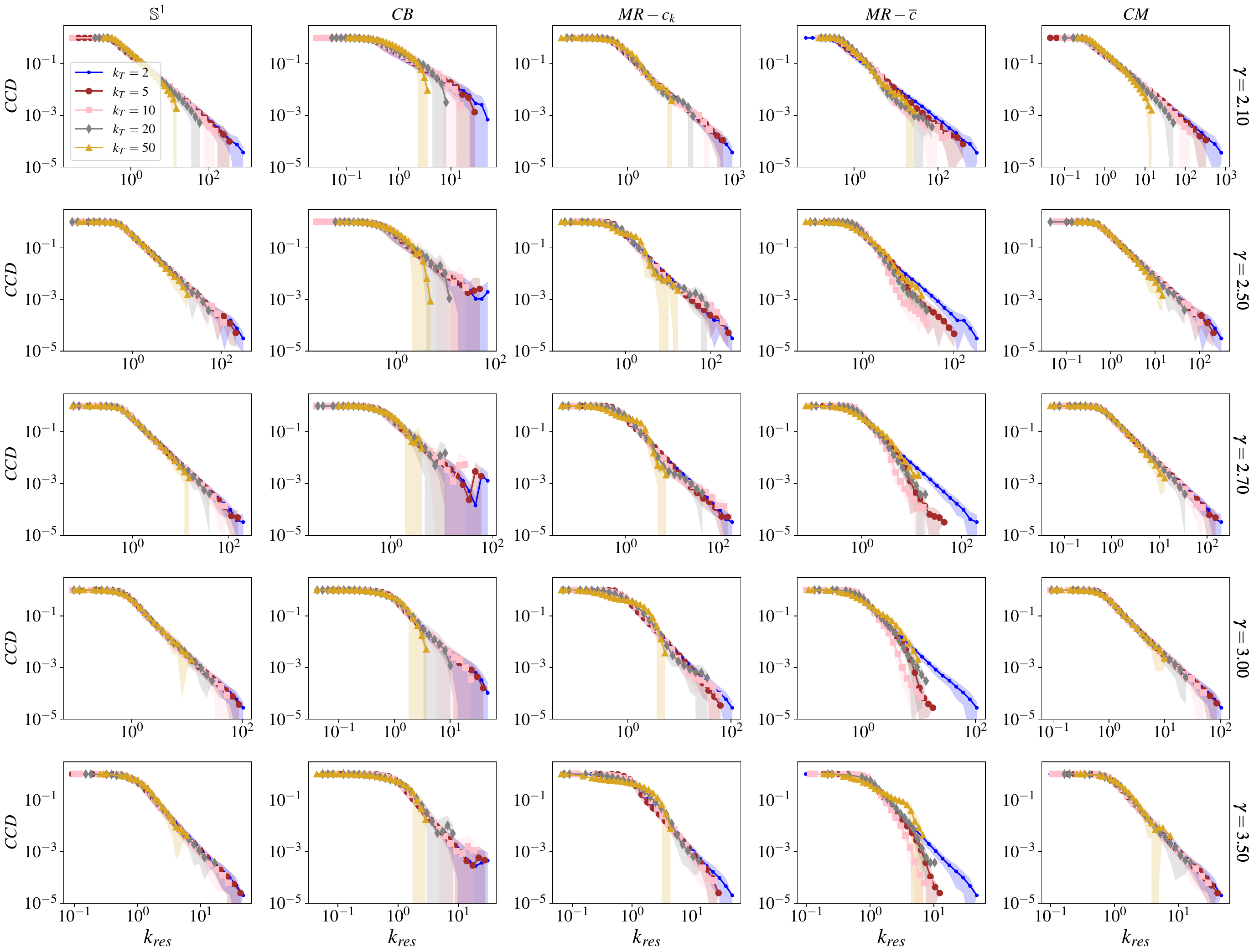}
\caption{The complementary cumulative degree distribution of the synthetic networks in the DTR flow as a function of the degree rescaled by the average degree in the corresponding subgraph, $k_{\text{res}} = k / \langle k(k_T) \rangle$. Each column corresponds to synthetic networks generated by a model with $N = 50000$ nodes, $\langle k \rangle = 10$, $\beta = 3$ and varying values of $\gamma$. The shaded areas represent two-$\sigma$ intervals around the mean, calculated from 10 realizations of each model. }\label{fig:Synthetic_CCD_b_3}
\end{figure*}

\begin{figure*}[t!]
\centering
\includegraphics[width=\textwidth]{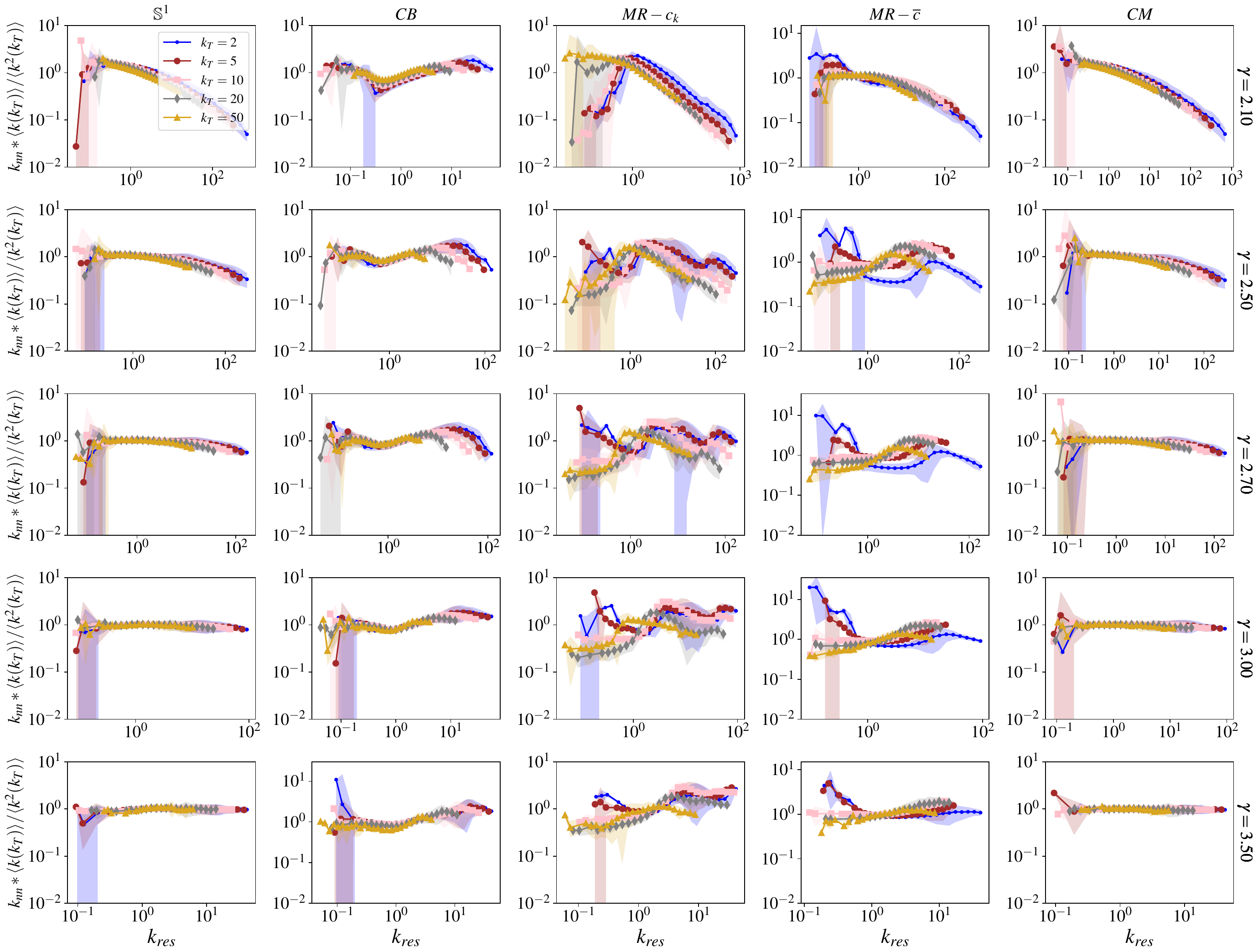}
\caption{Average nearest-neighbor degree of the synthetic networks in the DTR flow as a function of the degree rescaled by the average degree in the corresponding subgraph, $k_{\text{res}} = k / \langle k(k_T) \rangle$.  Each column corresponds to synthetic networks generated by a model with $N = 50000$ nodes, $\langle k \rangle = 10$, $\beta = 1.5$ and varying values of $\gamma$. The shaded areas represent two-$\sigma$ intervals around the mean, calculated from 10 realizations of each model.}\label{fig:Synthetic_Knn_b_1.5}
\end{figure*}

\begin{figure*}[t!]
\centering
\includegraphics[width=\textwidth]{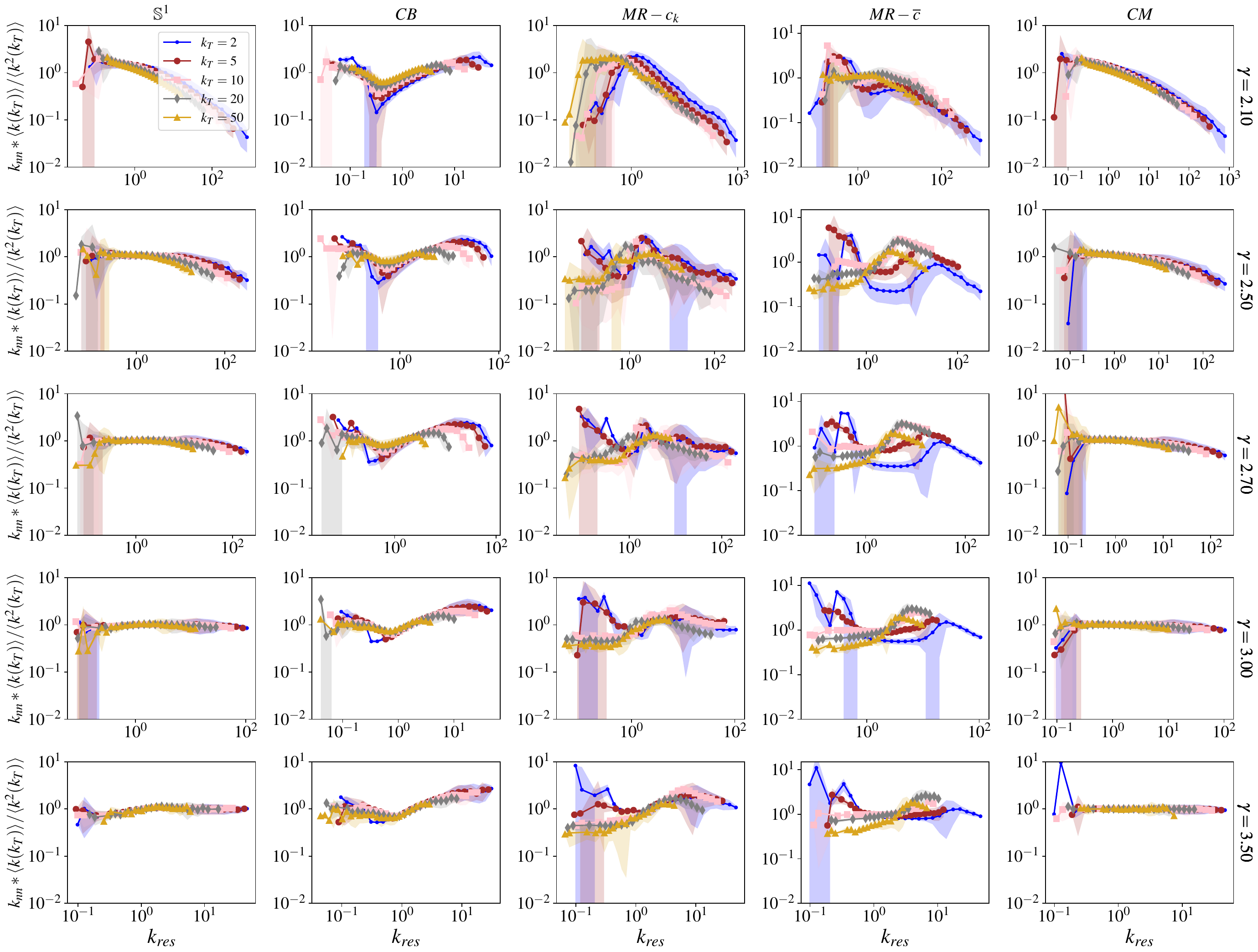}
\caption{Average nearest-neighbor degree of the synthetic networks in the DTR flow as a function of the degree rescaled by the average degree in the corresponding subgraph, $k_{\text{res}} = k / \langle k(k_T) \rangle$.  Eeach column corresponds to synthetic networks generated by a model with $N = 50000$, $\langle k \rangle = 10$, $\beta = 3$ and varying values of $\gamma$. The shaded areas represent two-$\sigma$ intervals around the mean, calculated from 10 realizations of each model.}\label{fig:Synthetic_Knn_b_3}
\end{figure*}

\begin{figure*}[t!]
\centering
\includegraphics[width=\textwidth]{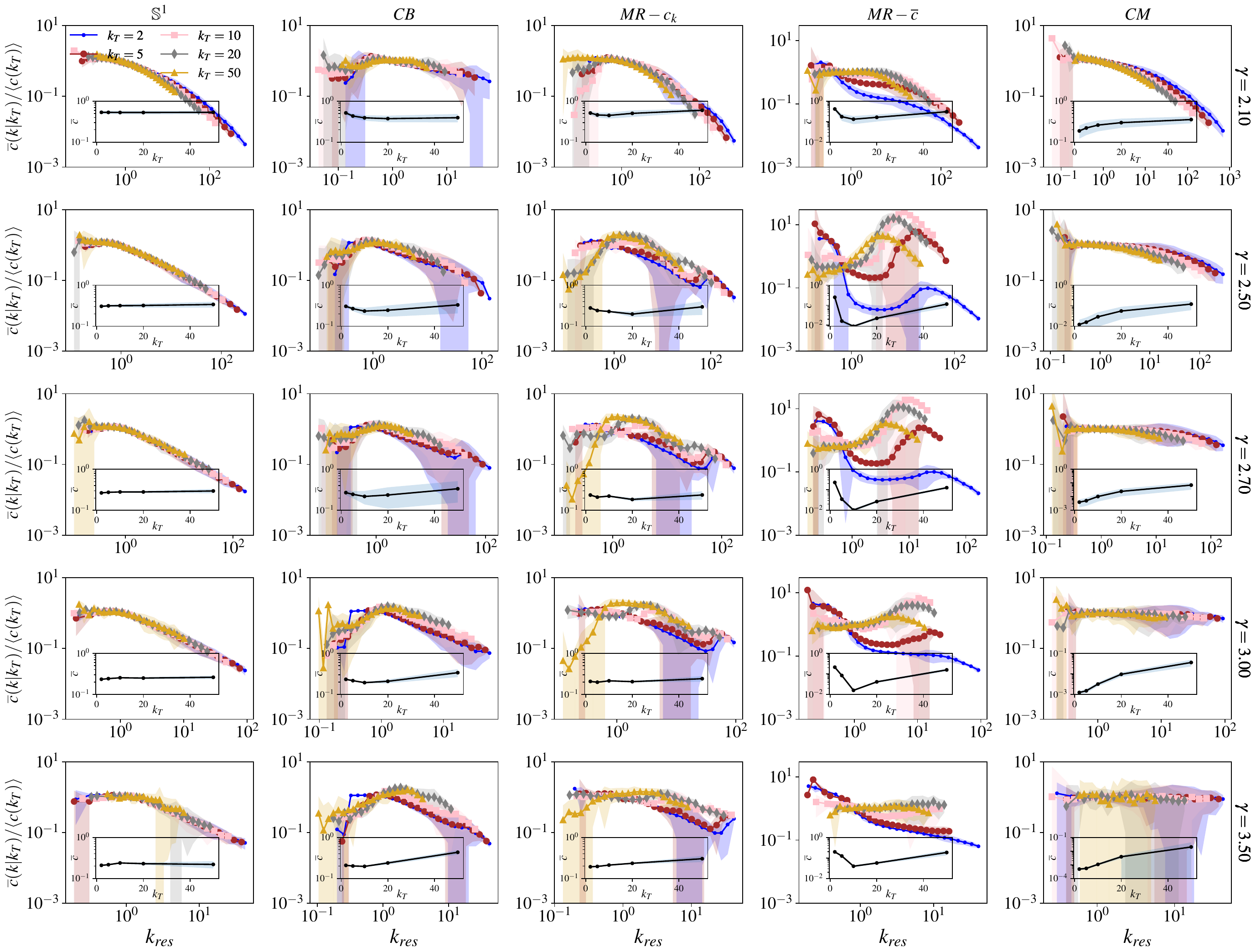}
\caption{Degree-dependent clustering coefficient of the synthetic networks in the DTR flow as a function of the degree rescaled by the average degree in the corresponding subgraph, $k_{\text{res}} = k / \langle k(k_T) \rangle$. Each column corresponds to synthetic networks generated by a model with $N = 50000$, $\langle k \rangle = 10$, $\beta = 1.5$ and varying values of $\gamma$. The inset highlights average clustering coefficient as a fucntion of $k_T$. The shaded areas represent two-$\sigma$ intervals around the mean, calculated from 10 realizations of each model.}\label{fig:Synthetic_clustering_b_1.5}
\end{figure*}

\begin{figure*}[t!]
\centering
\includegraphics[width=\textwidth]{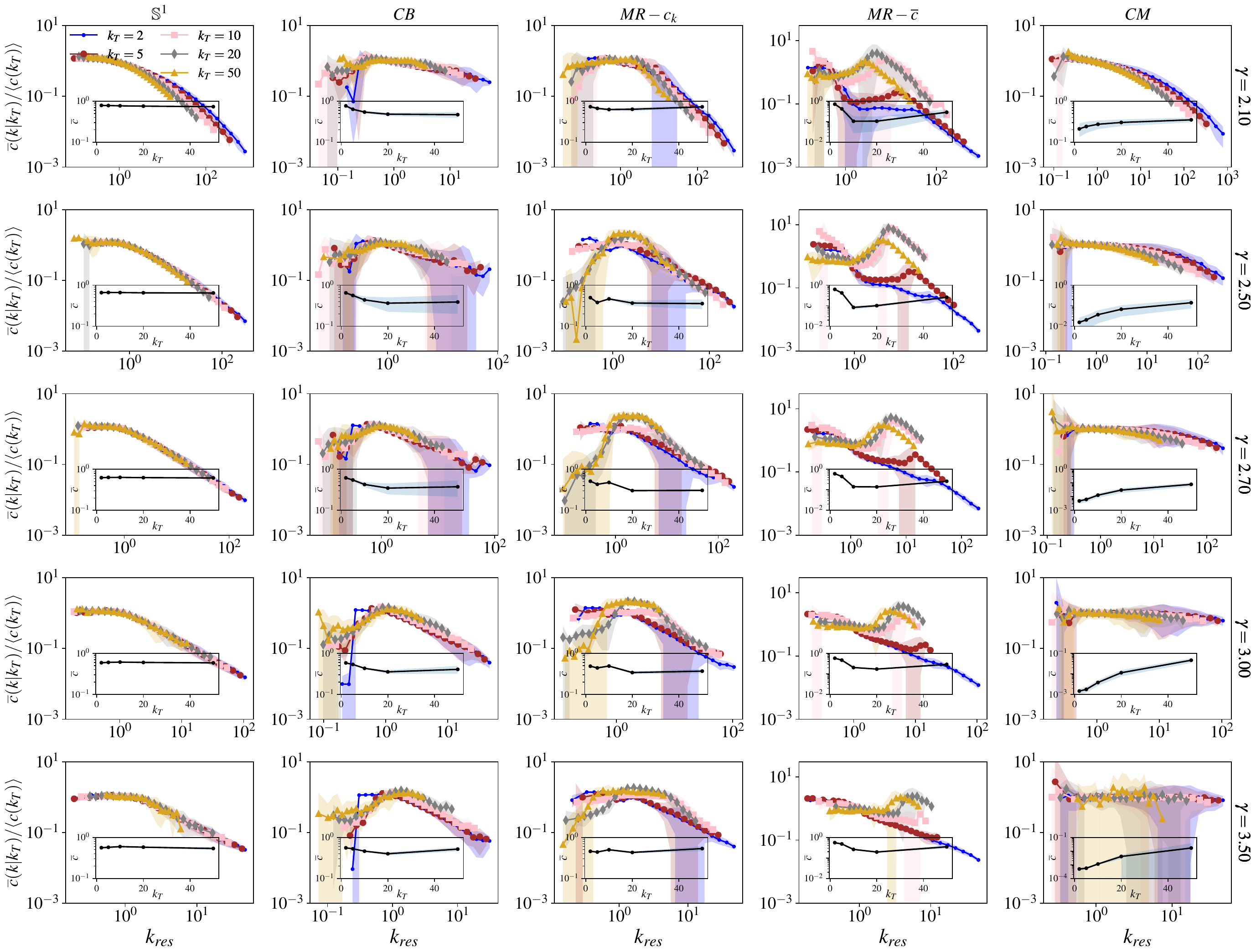}
\caption{Degree-dependent clustering coefficient of the synthetic networks in the DTR flow as a function of the degree rescaled by the average degree in the corresponding subgraph, $k_{\text{res}} = k / \langle k(k_T) \rangle$. Each column corresponds to synthetic networks generated by a model with $N = 50000$, $\langle k \rangle = 10$, $\beta = 3$ and varying values of $\gamma$. The inset highlights average clustering coefficient as a fucntion of $k_T$. The shaded areas represent two-$\sigma$ intervals around the mean, calculated from 10 realizations of each model.}\label{fig:Synthetic_clustering_b_3}
\end{figure*}

Figs.~\ref{fig:Synthetic_CCD_b_1.5} and \ref{fig:Synthetic_CCD_b_3} show the complementary cumulative degree distribution as a function of the rescaled degree $k_{res}$ in DTR flows for synthetic networks generated by the aforementioned models, with $\beta$ set to 1.5 and 3, respectively. Figs.~\ref{fig:Synthetic_Knn_b_1.5} and \ref{fig:Synthetic_Knn_b_3} highlight the average nearest-neighbor degree as a function of rescaled degree. Finally, Figs.~\ref{fig:Synthetic_clustering_b_1.5} and \ref{fig:Synthetic_clustering_b_3} present the clustering spectrum as a function of rescaled degree, with the insets showing the average clustering coefficient as a function of $k_T$. It is evident from these figures, particularly Figs.~\ref{fig:Synthetic_clustering_b_1.5} and \ref{fig:Synthetic_clustering_b_3}, that the $\mathbb{S}^1$ and $CM$ models exhibit self-similarity, the $CB$ model demonstrates quasi-self-similarity, and both the $MR-c_k$ and $MR-\overline{c}$ models are non-self-similar.

\section{$\epsilon^2$-test}  \label{sec:epsilon-test}
In order to classify networks as self-similar, quasi-self-similar, or non-self-similar, we measure variations in their structural properties in the DTR flows. In DTR, nodes with degree $k \leq k_T$ are removed from the given network to generate a hierarchy of its subgraphs. We then quantify the variations in the fundamental structural properties in the DTR flows, including the complementary cumulative degree distribution (CCD), the average nearest-neighbor degree function ($k_{nn}$), and the clustering spectrum ($\overline{c}(k)$).

Algorithm~\ref{AL:algo1} outlines the steps of the $\epsilon^2$-test, which computes $\epsilon^2$ values as the average squared relative difference between the curves representing structural properties of the original and renormalized networks. Based on the maximum value of $\epsilon^2$ ($\epsilon^2_{\text{Max}}$), the test categorizes networks into three types: networks with smaller $\epsilon^2_{\text{Max}}$ are classified as self-similar, moderate values indicate quasi-self-similarity, and larger values suggest non-self-similarity. The threshold values for this classification networks are determined by analyzing $\epsilon^2_{Max}$ values computed for a variety of real-world networks, as shown in Fig.~\ref{Fig:Real_Networks}-(b). These values are sorted in ascending order and plotted, forming a curve that reveals three distinct regions: low values represent self-similar networks, which exhibit a gradual increase with a very smooth slope, a moderate increase marks quasi-self-similar networks, and a sharp rise indicates non-self-similar networks. Based on this, we select $\epsilon^2_{Max} < 0.4$ for self-similar networks, $0.4 \leq \epsilon^2_{Max} < 1$ for quasi-self-similar networks, and $\epsilon^2_{Max} \geq 1$ for non-self-similar networks.  

Here, we consider the subgraph corresponding to $k_T = 2$ as the original network. Since low-degree nodes often introduce unwanted fluctuations in structural properties, removing them allows the analysis to focus on the core topological characteristics of the network. 

\begin{figure*}[t!]
\centering
\includegraphics[width=\textwidth]{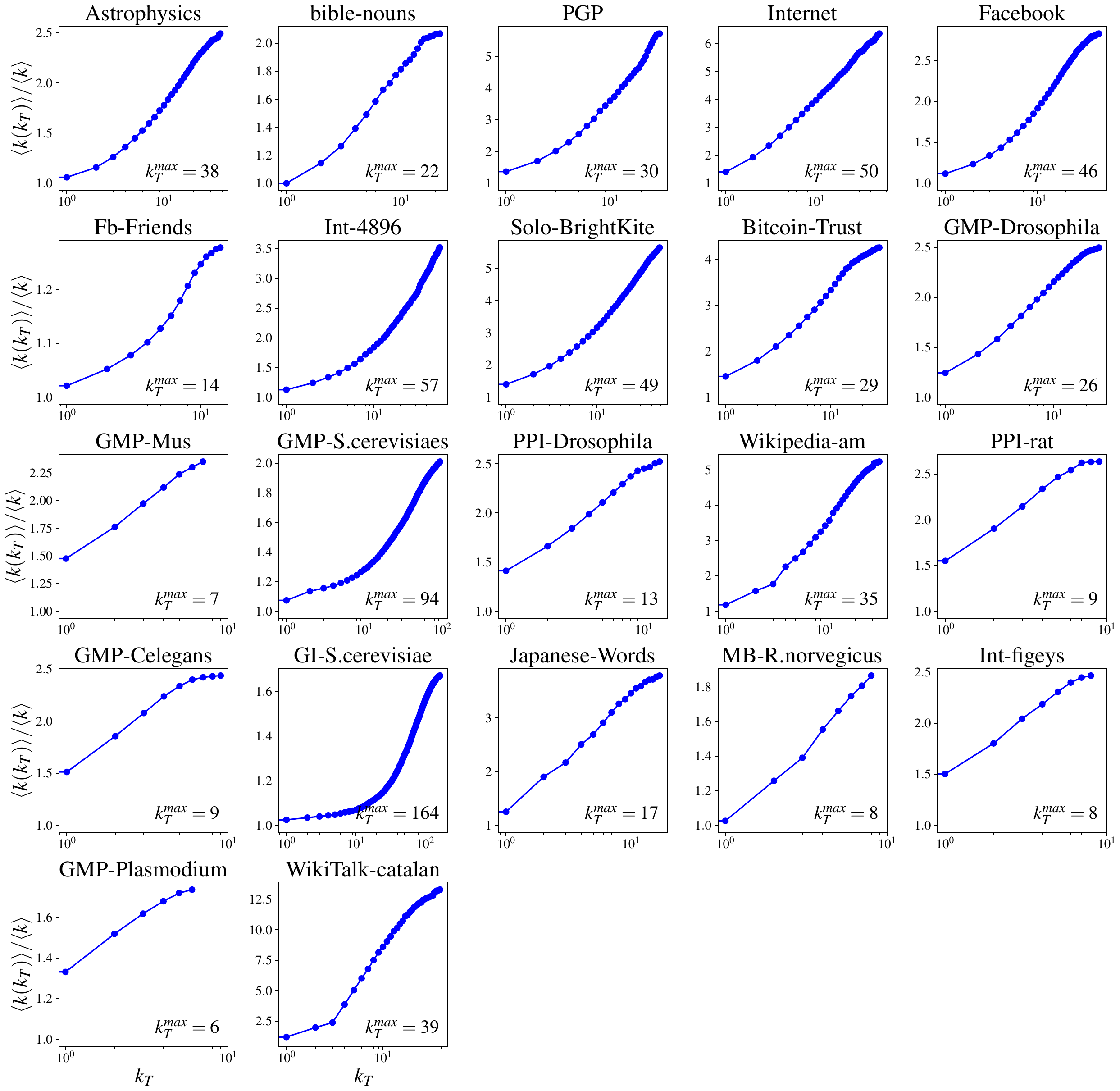}
\caption{ Average degree of subgraphs, normalized by the average degree of the original network as a function of $k_T$ in DTR flows. The maximum value for $k_T$ is the one at the peak of the curve.}\label{fig:Avg_Degree_vs_KT}
\end{figure*}

Fig.~\ref{fig:Avg_Degree_vs_KT} illustrates the normalized average degree of subgraphs, $\langle k(k_T) \rangle / \langle k \rangle$, as a function of $k_T$ in the DTR flows for real networks. The peak of this function is used to determine the $k_T^{max}$ value in real networks, accounting for finite-size effects and diverse degree distributions. Moreover, Table~\ref{tab:network_epsilon} highlights $\epsilon^2$ values computed for different structural properties in real networks.

In the experiments with real networks, the maximum number of intervals in $[2, k_T^{\text{max}}]$ is set to $n^{\text{max}} = 5$, the minimum number to $n^{\text{min}} = 2$ and the minimum gap between $k_T$ values is $\Delta k^{\text{min}} = 3$. The number of bins for the exponential binning is $n_{\text{bins}} = 20$ for both synthetic and real networks. 

\begin{algorithm}[H]

\caption{Algorithm for classifying networks as self-similar, quasi-self-similar, or non-self-similar}\label{AL:algo1}
\begin{algorithmic}[1]
\State Calculate the maximum value of the degree-thresholding parameter, $k_T^{\text{max}}$, corresponding to the peak of the curve $\langle k(k_T) \rangle / \langle k \rangle$ as a function of $k_T$.

\State Divide $[2, k_T^{\text{max}}]$ into $n$ evenly spaced intervals (except for the last one) to determine $n + 1$ values for $k_T$. Set $n = \max\big(n^{\text{min}}, \min\big(n^{\text{max}}, \lfloor (k_T^{\text{max}} - 2) / \Delta k \rfloor\big)\big)$, with $\Delta k = \max\big(\Delta k^{min}, \lfloor (k_T^{\text{max}} - 2) / n^{\text{max}} \rfloor \big)$.

\State For each value of $k_T$, apply DTR on the original network 
\State Plot $CCD$ as a function of rescaled degree, $k_{res} = k / \langle k (k_T) \rangle$, in the DTR flow
\State Plot $k_{nn} \times \langle k(k_T) \rangle / \langle k^2 (k_T) \rangle$ as a function of rescaled degree in the DTR flow
\State Plot $\overline{c}(k|k_T) / \langle c(k_T) \rangle$ as a function of rescaled degree in the DTR flow
\State Compute the average of the squared relative differences between each curve in the DTR flow and the one for original network across all three plots ($k_{nn}$, $CCD$, and \(\overline{c}(k)\)), using exponential binning with \( n_{\text{bins}} \) number of bins within the shared range of the x-axis for the two curves.

\[ \epsilon^2 = \frac{1}{n + 1} \frac{1}{n_{bins}} \sum_{k_T} \sum_{i=1}^{n_{bins}} \left( \frac{\text{bin value}^{org}_i - \text{bin value}^{k_T}_i}{\text{bin value}^{org}_i} \right)^2 \]
\State Select $\epsilon^2_{Max} = Max(\epsilon^2_{CCD}, \epsilon^2_{knn} , \epsilon^2_{\overline{c}})$
\If{$\epsilon ^2_{Max}< 0.4$}
	\State Classify the network as self-similar 
\ElsIf {$0.4 \leq \epsilon^2_{Max} <1$}
	\State Classify the network as Quasi-self-similar 
\Else
	\State Classify the network as Non-self-similar  
\EndIf

\end{algorithmic}
\end{algorithm}

\begin{table*}[htbp]
    \centering
    \caption{$\epsilon^2$ values computed for different structural properties including complementary comulative degree distribution $\epsilon^2_{CCD}$, average nearest-neighbor degree function $\epsilon^2_{k_{nn}}$, and clustering spectrum $\epsilon^2_{\overline{c}}$ in real-world network. The bolded values represent the maximum $\epsilon^2$ for each network, denoted as $\epsilon^2_{Max}$ and used to classify the networks into self-similar, quasi-self-similar, or non-self-similar. 
}
    \label{tab:network_epsilon}
    \begin{tabularx}{\textwidth}{>{\raggedright\arraybackslash}p{5cm} @{\hspace{1cm}} *{3}{X} }
        \toprule
      
        \hline
        Networks & $\epsilon^2_{CCD}$ & $\epsilon^2_{k_{nn}}$ & $\epsilon^2_{\overline{c}}$ \\
        \midrule
        \hline
		Astrophysics   & \textbf{0.29}  & 0.06 &0.13  \\        
        Bible-Nouns   & 0.25  & 0.16  & \textbf{0.26} \\
        PGP   & \textbf{0.35} & 0.16  & 0.24 \\
        Internet   &\textbf{0.32}  & 0.21  & 0.22 \\
        Facebook   & \textbf{0.22} & 0.07  &0.13  \\
        Fb-Friends   &0.12  & 0.05 & \textbf{0.13} \\
        Int-4896   & \textbf{0.29}  & 0.22  & 0.24  \\
        Solo-BrightKite   & 0.25  & 0.07  & \textbf{0.39}  \\
        Bitcoin-Trust   & \textbf{0.32} & 0.18  & 0.26  \\
        GMP-Drosophila   & 0.26  & 0.11  &  \textbf{0.59} \\
        GMP-Mus   &0.27  & 0.23  & \textbf{0.47} \\
        GMP–S.cerevisiae &\textbf{0.26}  & 0.08 &0.08  \\
        PPI-Drosophila  & 0.22  & 0.12  & \textbf{0.32}\\
        Wikipedia-am   &\textbf{0.38}  & 0.37  &0.33  \\
        PPI-rat   & 0.31 & 0.55 & \textbf{6.27}  \\
        GMP-Celegans  & 0.24  & 0.28  & \textbf{2.31}  \\
        GI-S.cerevisiae  & \textbf{0.27}  &0.05  &0.05  \\
        Japanese-Words   & \textbf{0.3}  & 0.21 & 0.23 \\
        GMP-Plasmodium   &\textbf{0.28}  &0.27  &0.26  \\
        MB-R.norvegicus   &\textbf{0.49}  & 0.45 & 0.48 \\
        Int-figeys   &0.24  & 0.33 & \textbf{1.13}  \\
        WikiTalk-Catalan  & \textbf{9.89} & 0.53  & 0.29 \\

        \hline
        \hline
    \end{tabularx}
\end{table*}


\section{Dataset description} \label{sec:Data-description}

\textbf{Astrophysics}~\cite{konect:newman01}: Nodes represent authors from the astrophysics section of arXiv (astro-p), and edges indicate collaborations.

\textbf{Bible-Nouns}~\cite{RomhildHarrisonBible}: Nodes are noun phrases (places and names) from the King James Bible, with edges denoting co-occurrence in verses.

\textbf{PGP}~\cite{PhysRevE.70.056122}: Nodes are users of the Pretty Good Privacy algorithm, and edges represent trust relationships through key signing.
    
\textbf{Internet}~\cite{Claffy4804445, Boguna2010}: Nodes are Autonomous Systems (ASs), and edges represent the connections between them in the network.

\textbf{Facebook}~\cite{10.1093/comnet/cnab014}: Nodes represent Facebook pages, and edges denote mutual likes between them.

\textbf{Fb-Friends}~\cite{Sapiezynski2019}: Nodes are university students from the Copenhagen Networks Study, and edges represent Facebook friendships.

\textbf{Int-4896}~\cite{doi:10.1073/pnas.1818013116}: Nodes represent the proteins of the Schizosaccharomyces pombe species, and edges denote experimentally verified physical interactions, including protein–protein, protein–DNA, metabolic pathway, and kinase–substrate interactions.

\textbf{Solo-BrightKite}~\cite{10.1145/2020408.2020579}: Nodes represent users of the Brightkite location-based social networking service, and edges indicate mutual friendships between users.

\textbf{Bitcoin-Trust}~\cite{kumar2016edge, kumar2018rev2}: Nodes represent users of the Bitcoin OTC platform, and directed edges indicate trust ratings, where each edge includes a weight assigned by one user to another. 

\textbf{GMP-Drosophila, GMP-Mus, GMP–S.cerevisiae, GMP-Celegans, GMP-Plasmodium}~\cite{10.1093/comnet/cnu038}: Multiplex networks representing different types of genetic interactions in various organisms. Nodes represent genes, and edges capture different genetic interactions across different layers, including physical, association, co-localization, direct, suppressive, and additive/synthetic interactions.  

\textbf{PPI-Drosophila}~\cite{Tang2023}: Nodes represent proteins in the Drosophila melanogaster species, and edges represent binary physical protein-protein interactions identified using yeast two-hybrid (Y2H) analysis.

\textbf{Wikipedia-am}~\cite{kunegis2013konect}: Nodes represent articles in the Amharic language edition of Wikipedia, and directed edges represent hyperlinks between articles

\textbf{PPI-rat}~\cite{10.1093/nar/gkx1116}: Nodes represent proteins in the rat species, and edges represent physical protein-protein interactions between them.

\textbf{GI-S.cerevisiae}~\cite{10.1093/nar/gkx1116}: Nodes represent genes in Saccharomyces cerevisiae (baker's yeast), and edges indicate genetic interactions where mutations in one gene affect or are modified by mutations in another gene.

\textbf{Japanese-Words}~\cite{doi:10.1126/science.1089167}: Nodes represent words in Japanese texts, and edges indicate adjacency, where one word directly follows another. The original network is directed.

\textbf{MB-R.norvegicus}~\cite{huss2007currency}: Nodes represent substances involved in enzymatic reactions in Rattus norvegicus (rat), and edges indicate reactant-product relationships in these reactions.

\textbf{Int-figeys}~\cite{Ewing2007}: Nodes represent human proteins, and edges denote binding interactions between proteins.

\textbf{WikiTalk-Catalan}~\cite{kunegis2013konect}: Nodes represent registered editors in the Catalan Wikipedia, and edges indicate interactions where one user writes a message on another user's talk page.

In this work, we ignore weights in the weighted networks and construct undirected versions of all directed networks. In the case of multiplex networks, we create a monolayer network by treating all interaction types equally and removing repeated links.

\begin{table*}[htbp]
    \centering
    \caption{Real network Properties. $N$ is the number of nodes, $E$ is the number of edges, $\langle k \rangle$ is average degree, $k_{\text{max}}$ is maximum degree and $\overline{c}$ is the average clustering in the giant connected component of the real networks. $\beta$ indicates the level of clustering and computed using Mercator embedding tool. Type S refers to self-similar networks, Type NS indicates non-self-similar networks, and Type QS denotes quasi-self-similar networks. The following abbreviations are used: (MB) for metabolic, (GI) for genetic interactions, (GMP) for genetic multiplex, (PPI) for protein-protein interactions, (Solo) for social locations, and (Int) for interactome. 
     }
    \label{tab:network_properties}
   \begin{tabularx}{\textwidth}{>{\raggedright\arraybackslash}p{4.3cm} *{5}{X} >{\raggedright\arraybackslash}p{1.3cm} *{2}{X}}

        \toprule
        \hline

        Networks & $N$ & $E$ & $\langle k \rangle$ & $k_{\text{max}}$ & $\overline{c}$ & $\beta$  & $\epsilon^2_{Max}$ & Type \\
        \midrule
        \hline
        
        Astrophysics  & 14845      & 119652     & 16.12   & 360     & 0.72   & 3.27    & 0.29   & S      \\        
        
        Bible-Nouns    & 1707  & 9059  & 10.61  & 364  & 0.71  &  3.2  &  0.26  & S   \\        
        
        PGP    & 10680     & 24316    & 4.55   & 205     & 0.44    & 1.95  &  0.35   & S     \\         

      	Internet    & 23748   & 58414    & 4.92    & 2778  &0.61    & 1.95  & 0.32    &  S      \\

		Facebook    & 22470     & 170823     & 15.2     & 709     & 0.41    & 1.6   & 0.22    & S    \\

		Fb-Friends   & 800 & 6418 & 16.05 &101  & 0.32 & 1.54 &   0.12   & S  \\ 

		Int-4896  & 4086  &47961   & 23.48   & 448    &0.41    & 1.43   &  0.29   &  S  \\

        Solo-BrightKite  & 56739  & 212945    & 7.51   &1134   &0.27   & 1.33  & 0.39 & S   \\         
        
        Bitcoin-Trust  &5875   & 21489   & 7.31   & 795  & 0.29 & 1.13   & 0.32  &  S   \\        
		
		GMP-Drosophila   & 8114    &38909    &9.6   & 179   &0.12    &  1.07  & 0.59  &  QS   \\        
        
		GMP-Mus   &7402   &16858    & 4.56    & 368   & 0.13    &  0.99  & 0.47   & QS   \\        
        
        GMP–S.cerevisiae & 6567  & 223539  & 68.08   & 3254   &0.22   & 0.87  & 0.26   & S   \\        

        PPI-Drosophila  &2705   & 8458  & 6.25  & 129   &0.07   & 0.78  & 0.32   &  S   \\

		Wikipedia-am  & 20883  & 94022  & 9   & 3911  & 0.18  &  0.76 & 0.38  &  S  \\        
        
		PPI-rat  & 6803 & 14636   & 4.3     & 836   & 0.15    &  0.72  & 6.27  & NS      \\        
        
		GMP-Celegans  &3692    & 7650    &4.14     & 526    & 0.11   & 0.69   & 2.31  & NS    \\        
        
        GI-S.cerevisiae  &5933     & 441991    & 148.99    &  2244   & 0.17    & 0.63  & 0.27  & S     \\                  
		Japanese-Words  &2698   &7995   &5.93     &725     &0.3     &  $\beta \approx 0$   & 0.3  &S    \\
        
        		GMP-Plasmodium  &1158    & 2402   & 4.15    & 83    &0.03     &  $\beta \approx 0$  & 0.28  & S  \\
        		
		MB-R.norvegicus   & 1590  & 4666   & 5.87    & 498  & 0.19  &  $\beta \approx 0$  & 0.49  & QS   \\ 
		
		Int-figeys  & 2217   & 6418   & 5.79   & 314      & 0.07    & $\beta \approx 0$   & 1.13    & NS    \\ 
		
		WikiTalk-Catalan  & 79209  & 181529  & 4.59   & 53234    &0.83 &$\beta \approx 0$  &9.89 &NS   \\
               
        \hline

        \bottomrule
    \end{tabularx}
\end{table*}

\section{Experimental Analysis of Real-World Networks} \label{sec:Exp-Real}
\begin{figure*}[t!]
\centering
\includegraphics[width=\textwidth]{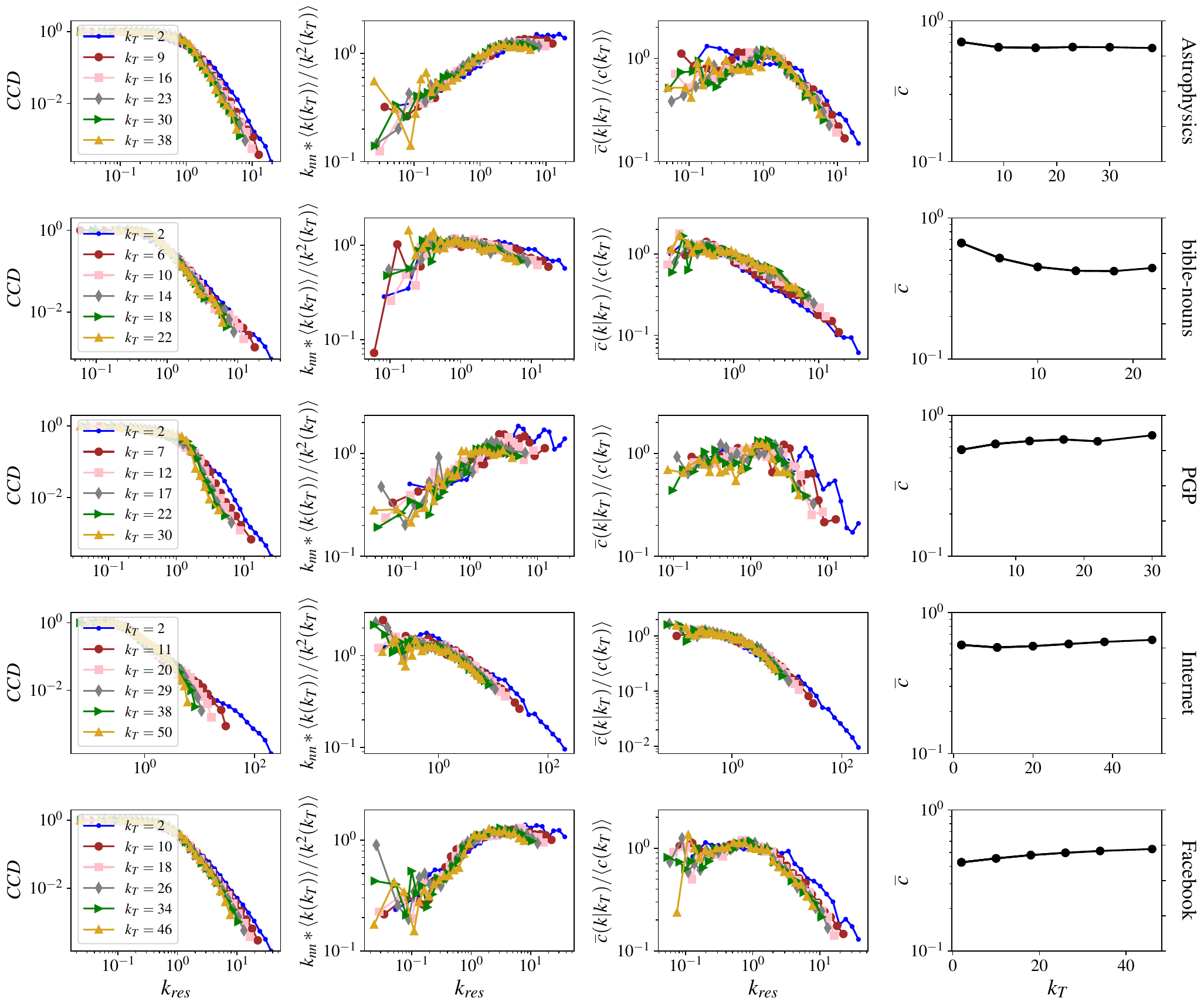}
\caption{Topological properties of real-world \textbf{self-similar} networks, including the complementary cumulative degree distribution (CCD), the rescaled average nearest-neighbor degree $k_{nn} * \langle k(k_T) \rangle / \langle k^2(k_T) \rangle$, the rescaled clustering spectrum $\overline{c} (k | k_T) / \langle c(k_T) \rangle$ as a function of the rescaled degree $k_{\text{res}} = k / \langle k(k_T) \rangle$, and the average clustering coefficient $\overline{c}$ as a function of $k_T$ in the DTR flow.}
\label{fig:Real_Net_Self_Similar1}
\end{figure*}

\begin{figure*}[t!]
\centering
\includegraphics[width=\textwidth]{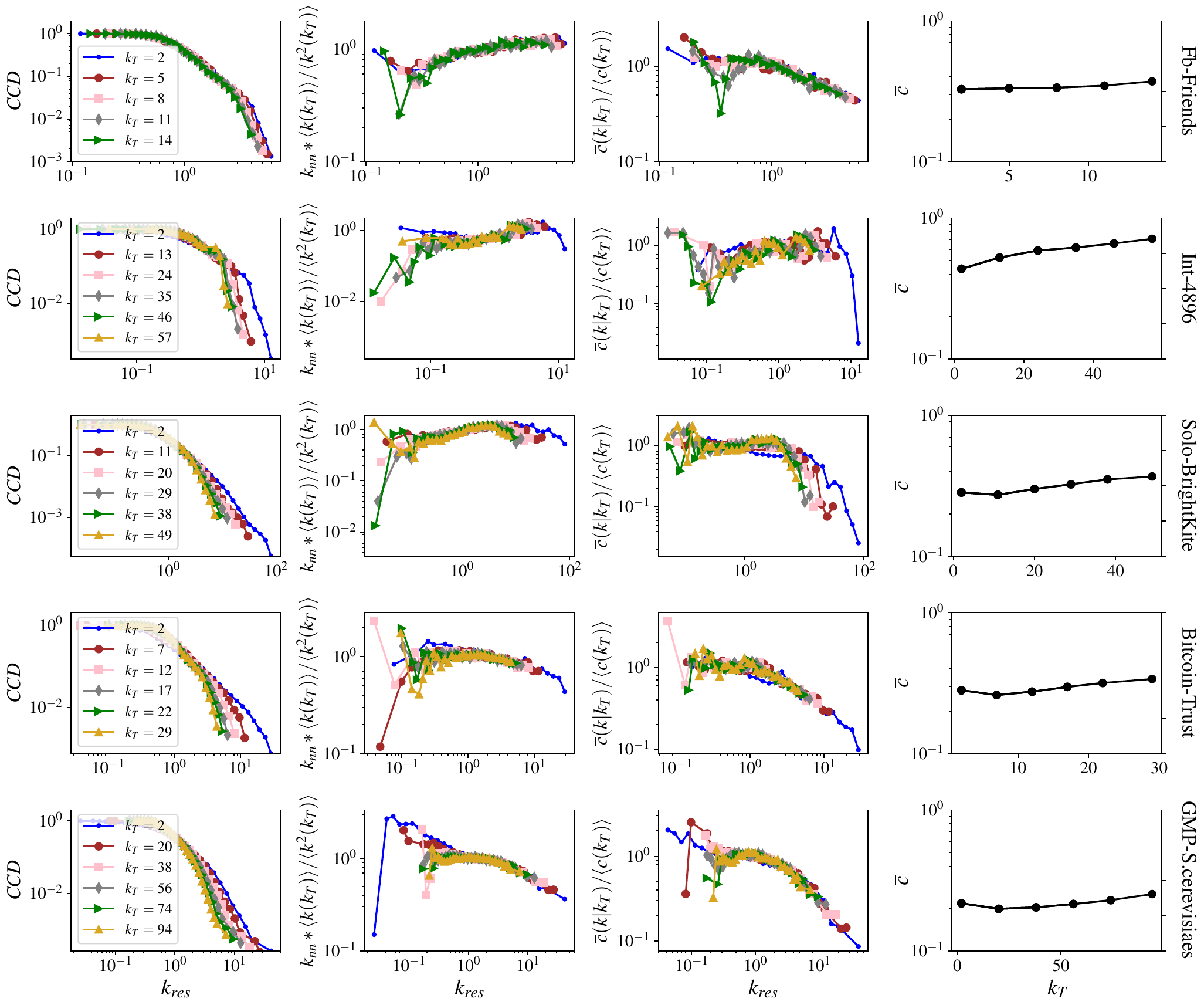}
\caption{Topological properties of real-world \textbf{self-similar} networks, including the complementary cumulative degree distribution (CCD), the rescaled average nearest-neighbor degree $k_{nn} * \langle k(k_T) \rangle / \langle k^2(k_T) \rangle$, the rescaled clustering spectrum $\overline{c} (k | k_T) / \langle c(k_T) \rangle$ as a function of the rescaled degree $k_{\text{res}} = k / \langle k(k_T) \rangle$, and the average clustering coefficient $\overline{c}$ as a function of $k_T$ in the DTR flow.}\label{fig:Real_Net_Self_Similar2}
\end{figure*}

\begin{figure*}[t!]
\centering
\includegraphics[width=\textwidth]{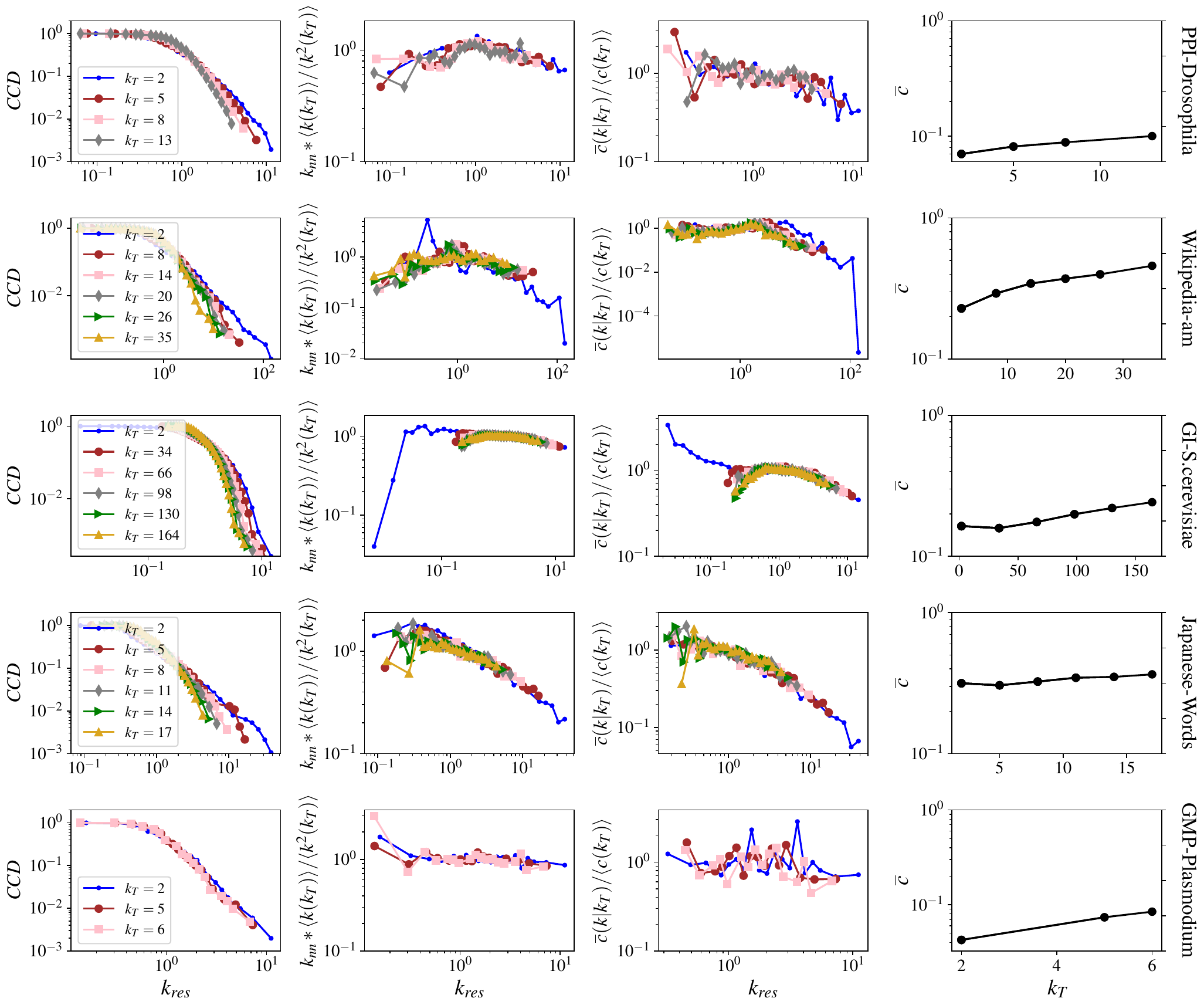}
\caption{Topological properties of real-world \textbf{self-similar} networks, including the complementary cumulative degree distribution (CCD), the rescaled average nearest-neighbor degree $k_{nn} * \langle k(k_T) \rangle / \langle k^2(k_T) \rangle$, the rescaled clustering spectrum $\overline{c} (k | k_T) / \langle c(k_T) \rangle$ as a function of the rescaled degree $k_{\text{res}} = k / \langle k(k_T) \rangle$, and the average clustering coefficient $\overline{c}$ as a function of $k_T$ in the DTR flow.}\label{fig:Real_Net_Self_Similar3}
\end{figure*}

\begin{figure*}[t!]
\centering
\includegraphics[width=\textwidth]{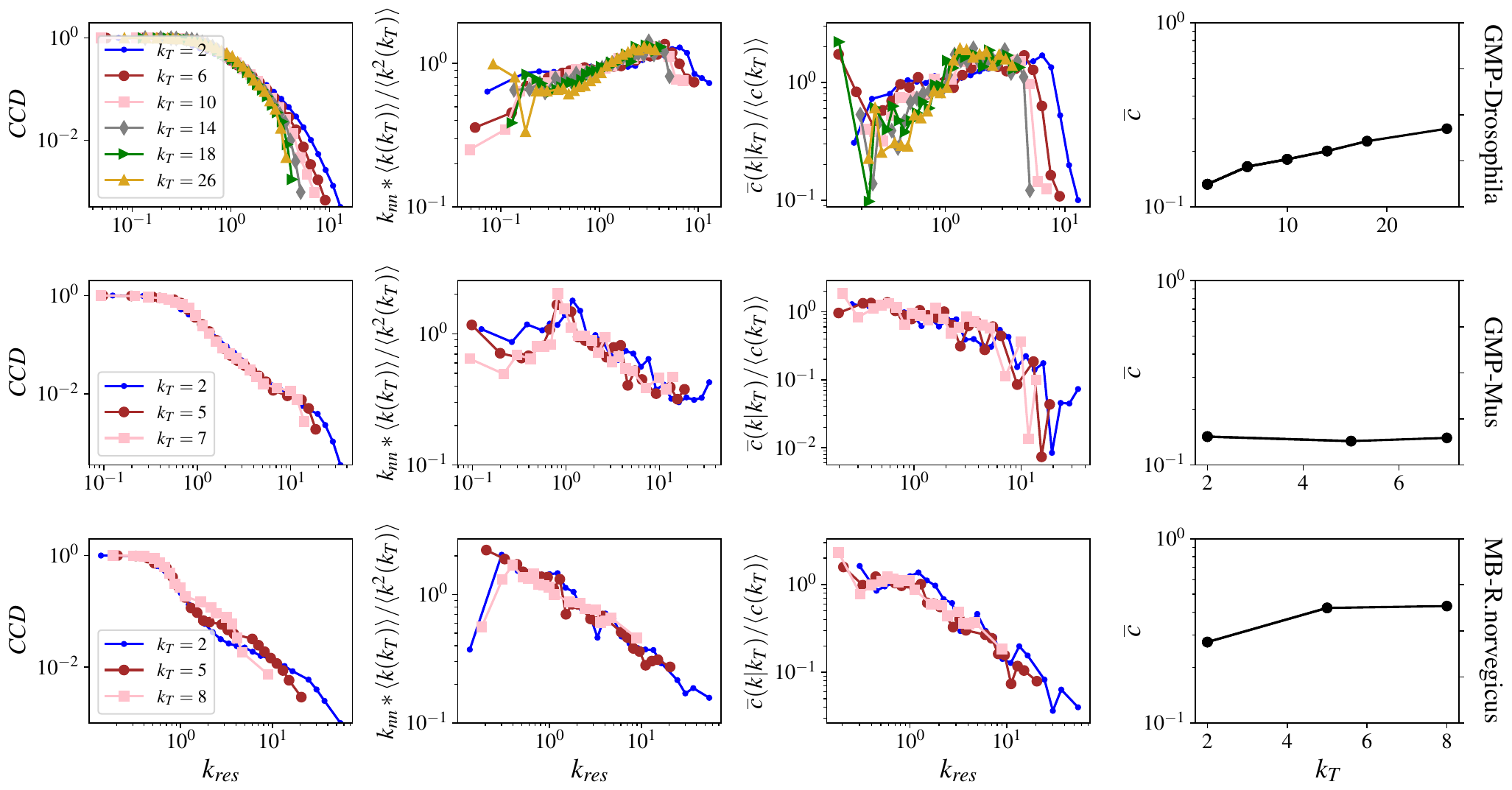}
\caption{Topological properties of real-world \textbf{quasi-self-similar} networks, including the complementary cumulative degree distribution (CCD), the rescaled average nearest-neighbor degree $k_{nn} * \langle k(k_T) \rangle / \langle k^2(k_T) \rangle$, the rescaled clustering spectrum $\overline{c} (k | k_T) / \langle c(k_T) \rangle$ as a function of the rescaled degree $k_{\text{res}} = k / \langle k(k_T) \rangle$, and the average clustering coefficient $\overline{c}$ as a function of $k_T$ in the DTR flow.}\label{fig:Real_Net_Quasi_Self_Similar}
\end{figure*}

\begin{figure*}[t!]
\centering
\includegraphics[width=\textwidth]{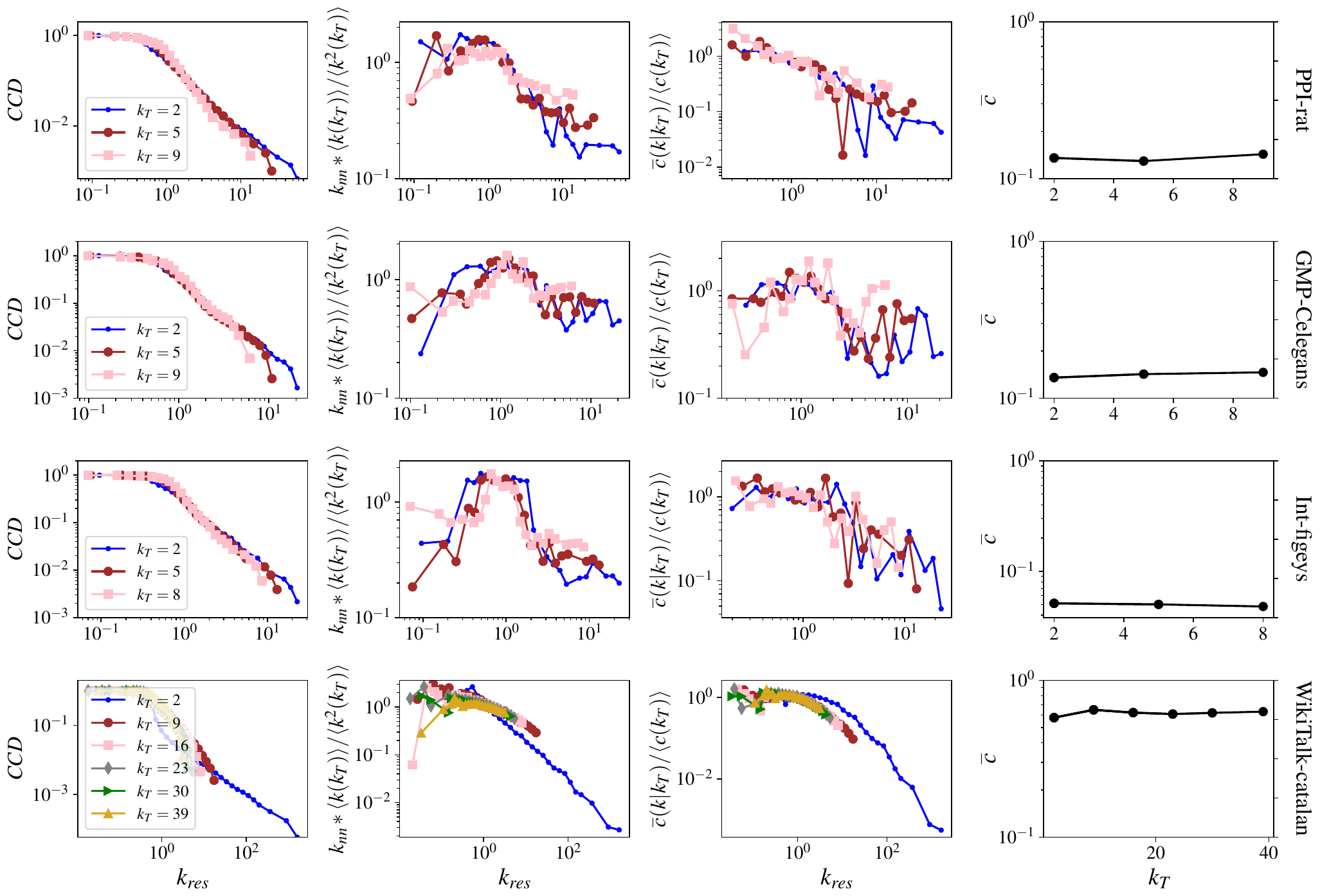}
\caption{Topological properties of real-world \textbf{non-self-similar} networks, including the complementary cumulative degree distribution (CCD), the rescaled average nearest-neighbor degree $k_{nn} * \langle k(k_T) \rangle / \langle k^2(k_T) \rangle$, the rescaled clustering spectrum $\overline{c} (k | k_T) / \langle c(k_T) \rangle$ as a function of the rescaled degree $k_{\text{res}} = k / \langle k(k_T) \rangle$, and the average clustering coefficient $\overline{c}$ as a function of $k_T$ in the DTR flow.}\label{fig:Real_Net_Non_Self_Similar}
\end{figure*}

Here, we analyze the DTR self-similarity in real networks. Table~\ref{tab:network_properties} summarizes the key properties of these networks.
Figs.~\ref{fig:Real_Net_Self_Similar1}, \ref{fig:Real_Net_Self_Similar2}, and \ref{fig:Real_Net_Self_Similar3} showcase how the fundamental structural properties of self-similar networks varies under DTR. As expected for self-similar networks and in agreement with the $\epsilon^2$-Test values reported in Table~\ref{tab:network_properties}, the degree distribution, average nearest-neighbor degree, and clustering spectrum exhibit minimal variation across different values of $k_T$. Similarly, Figs.~\ref{fig:Real_Net_Quasi_Self_Similar} and \ref{fig:Real_Net_Non_Self_Similar} depict the corresponding variations for quasi-self-similar and non-self-similar networks, respectively. These variations are more pronounced for quasi-self-similar and non-self-similar networks, highlighting their weaker structural invariance under DTR.

\begin{figure*}[t!]
\centering
\includegraphics[width=0.7\textwidth]{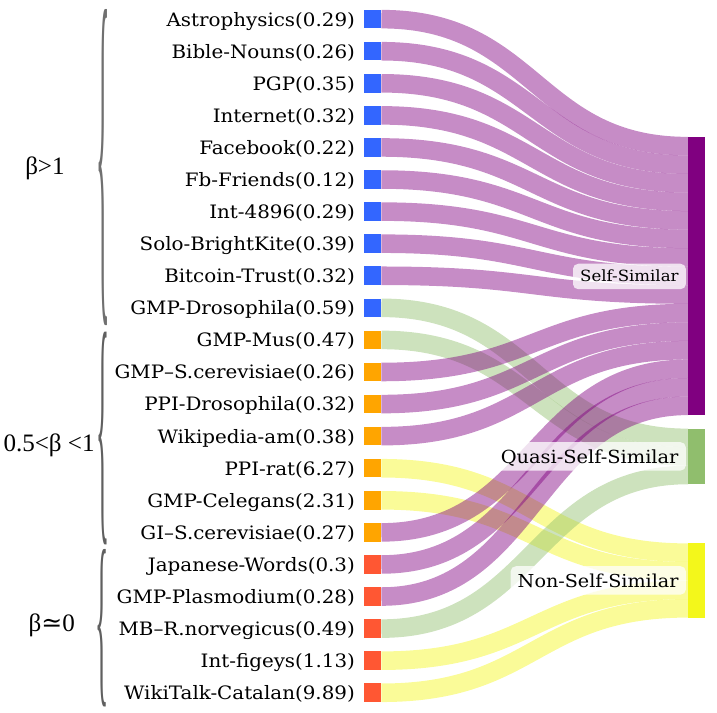}
\caption{Sankey diagram of real networks, sorted from the top in descending order by $\beta$ values. On the left, networks are classified as geometric ($\beta > 1$), quasi-geometric ($0.5 < \beta < 1$), and non-geometric ($\beta \approx 0$). On the right, they are categorized in terms of self-similarity, using their $\epsilon^2_{\text{Max}}$ values displayed in parentheses.}\label{fig:Sankey}
\end{figure*}

Moreover, Fig.~\ref{fig:Sankey} presents a Sankey diagram for the real-world networks sorted from the top in descending order of their estimated $\beta$, which quantifies the coupling between network topology and the underlying hyperbolic space of the $\mathbb{S}^1$ model.  On the left side, they are categorizing as self-similar, quasi-self-similar, or non-self-similar based on their $\epsilon^2$-Test values displayed in paranthesis and the thresholds from Fig.~\ref{Fig:Real_Networks}-(b). The results highlight that all non-self-similar networks have $\beta < 1$, classifying them as either non-geometric or quasi-geometric. Additionally, nearly all geometric networks are self-similar, with the exception of GMP-Drosophila, which has a $\beta$ value near the critical threshold $\beta_c = 1$ and is thus classified as quasi-self-similar. These findings reinforce the connection between self-similarity and geometricity in real networks.

\appendix

\clearpage
\clearpage
\bibliography{refs}

\end{document}